\documentclass[aps,pra,twocolumn,superscriptaddress,reprint]{revtex4-2}

\usepackage{graphicx}
\usepackage{dcolumn}
\usepackage{bm}
\usepackage[utf8]{inputenc}
\usepackage[T1]{fontenc}
\usepackage{mathptmx}
\usepackage{textgreek}
\usepackage{hyperref}
\usepackage{amsmath}
\usepackage{upgreek}
\usepackage{hhline}
\usepackage{textcomp,gensymb}
\usepackage{blindtext}
\usepackage[per=slash]{siunitx}
\usepackage{color}
\usepackage{nicefrac}
\usepackage{bm}
\usepackage[]{changes}
\usepackage{textgreek}
\usepackage{textcmds}

\DeclareSIUnit{\intensity}{\watt\per\centi\meter\squared}
\DeclareSIUnit{\electrondensity}{\textit{e}\per\centi\meter\cubed}

\begin{document}

	\title{Characterization of 1- and 2-\textmu m-wavelength laser-produced microdroplet-tin plasma \\ for generating extreme-ultraviolet light}
	\author{R Schupp}
	\affiliation{Advanced Research Center for Nanolithography, Science Park~106, 1098~XG Amsterdam, The Netherlands}
	\author{L Behnke}
	\affiliation{Advanced Research Center for Nanolithography, Science Park~106, 1098~XG Amsterdam, The Netherlands}
	\author{J Sheil}
	\affiliation{Advanced Research Center for Nanolithography, Science Park~106, 1098~XG Amsterdam, The Netherlands}
	\author{Z Bouza}
	\affiliation{Advanced Research Center for Nanolithography, Science Park~106, 1098~XG Amsterdam, The Netherlands}
	\author{M Bayraktar}
	\affiliation{Industrial Focus Group XUV Optics, MESA+ Institute for Nanotechnology, University of Twente, Drienerlolaan 5, 7522 NB Enschede, The Netherlands}
	\author{W Ubachs}
	\affiliation{Advanced Research Center for Nanolithography, Science Park~106, 1098~XG Amsterdam, The Netherlands}
	\affiliation{Department of Physics and Astronomy, and LaserLaB, Vrije Universiteit, De Boelelaan 1081, 1081 HV Amsterdam, The Netherlands}
	\author{R Hoekstra}
	\affiliation{Advanced Research Center for Nanolithography, Science Park~106, 1098~XG Amsterdam, The Netherlands}
	\affiliation{Zernike Institute for Advanced Materials, University of Groningen, Nijenborgh 4, 9747 AG Groningen, The Netherlands}
	\author{O O Versolato}\email{o.versolato@arcnl.nl}
	\affiliation{Advanced Research Center for Nanolithography, Science Park~106, 1098~XG Amsterdam, The Netherlands}
	\affiliation{Department of Physics and Astronomy, and LaserLaB, Vrije Universiteit, De Boelelaan 1081, 1081 HV Amsterdam, The Netherlands}
	
	\date{\today}
	
	\begin{abstract}
		Experimental spectroscopic studies are presented, in a 5.5--25.5\,nm extreme-ultraviolet (EUV) wavelength range, of the light emitted from plasma produced by the irradiation of tin microdroplets by 5-ns-pulsed, 2-\textmu m-wavelength laser light. 
		Emission spectra are compared to those obtained from plasma driven by 1-\textmu m-wavelength Nd:YAG laser light over a range of laser intensities spanning approximately $0.3 - 5 \times 10^{11}$\,W\,cm$^{-2}$, under otherwise identical conditions. Over this range of drive laser intensities, we find that similar spectra and underlying plasma charge state distributions are obtained when keeping the ratio of 1-\textmu m to 2-\textmu m laser intensities fixed at a value of 2.1(6), which is in good agreement with RALEF-2D radiation-hydrodynamic simulations. Our experimental findings, supported by the simulations, indicate an approximately inversely proportional scaling $\sim \lambda^{-1}$ of the relevant plasma electron density, and of the aforementioned required drive laser intensities, with drive laser wavelength $\lambda$. This scaling also extends to the optical depth that is captured in the observed changes in spectra over a range of droplet diameters spanning 16-51\,\textmu m at a constant laser intensity that maximizes the emission in a 2\% bandwidth around \SI{13.5}{nm} relative to the total spectral energy, the bandwidth relevant for EUV lithography.
		The significant improvement of the spectral performance of the 2-\textmu m- vs 1-\textmu m driven plasma provides strong motivation for the development of high-power, high-energy near-infrared lasers to enable the development of more efficient and powerful sources of EUV light.
	\end{abstract}

	\maketitle
	
    \section{Introduction}
	Laser-driven microdroplet-tin plasma provides the extreme-ultraviolet (EUV) light that is used in state-of-the-art EUV lithography\,\cite{versolato2019physics,Purvis2018industrialization,Moore2018euv,Schafgans2015performance,OSullivan2015,Banine2011,Benschop2008}. Ever more powerful sources of EUV light are required for future lithography applications. This EUV light is generated from electronic transitions in multiply-charged tin ions that strongly emit radiation in a narrow band around 13.5\,nm\,\cite{Azarov1993,Churilov2006SnIX--SnXII,Churilov2006SnVIII,Churilov2006SnXIII--XV,Ryabtsev2008SnXIV,Tolstikhina2006ATOMICDATA,DArcy2009a,Ohashi2010,Colgan2017,Torretti2017,Scheers2020}. EUV-emitting plasma in an industrial nanolithography machine is driven by CO$_2$-gas lasers with a 10-\textmu m wavelength. Such plasma achieves particularly high conversion efficiencies (CE) of converting drive laser light into EUV radiation in a 2-\% wavelength bandwidth around 13.5\,nm that can be transported by the available Mo/Si multilayer optics\,\cite{Bajt2002,Huang2017}. Near- or mid-infrared solid-state lasers may however soon become an attractive alternative to the CO$_2$-gas lasers because such modern solid-state lasers are expected to have a significantly higher efficiency in converting electrical power to laser light. Furthermore, they may reach much higher pulse energies and output powers, in turn enabling more EUV output. Big Aperture Thulium (BAT) lasers\,\cite{Danson2019petawatt,Sistrunk2019} represent a particularly promising class of novel, powerful laser systems that has recently drawn significant attention. These lasers would operate at 1.9-\textmu m wavelength, in between the well-known cases of 1- and 10-\textmu m drive lasers. Recent simulation work indicates that a global CE optimum lies within this range of 1- and 10-\textmu m drive laser wavelength\,\cite{Siders2019euvlitho}. Briefly, such studies point out that the longer-wavelength drivers are associated with sub-optimal absorption of the laser energy by the plasma whereas shorter-wavelength drivers may exhibit severe opacity broadening of the EUV spectrum out of the 2-\% acceptance bandwidth \cite{Schupp2019b,Freeman2012laser,Harilal2011effect}. To date, no experimental studies of mass-limited, microdroplet-tin-based plasmas driven by lasers in this wavelength range are however available to verify these claims.
	
	In this article a study of the EUV emission spectrum of 2-\textmu m-wavelength-laser-driven tin-microdroplet plasma is presented. The laser light is obtained from a master oscillator power amplifier setup that comprises a series of KTP crystals pumped by a ns-pulsed Nd:YAG laser ($\lambda=1$\,\textmu m), enabling to gauge the potential of, e.g., thulium lasers without the effort of building one. The recorded spectroscopic data are compared to those obtained from a 1-\textmu m-driven plasma under otherwise identical conditions, over a wide range of droplet sizes and laser intensities. Radiation-hydrodynamic simulations using the RALEF-2D code \cite{Basko2010development}, are presented to support the experimental findings. Following the recent work of Schupp \textit{et al}.\,\cite{Schupp2019b} on Nd:YAG-laser-pumped plasma, an analytical solution for radiation transport in an optically thick one-dimensional plasma is used to quantify the influence of optical depth on the broadening of the key emission feature at 13.5\,nm.

	\section{Experiment}
	
\begin{figure}[tb]
		\centering
		\includegraphics[width=\linewidth]{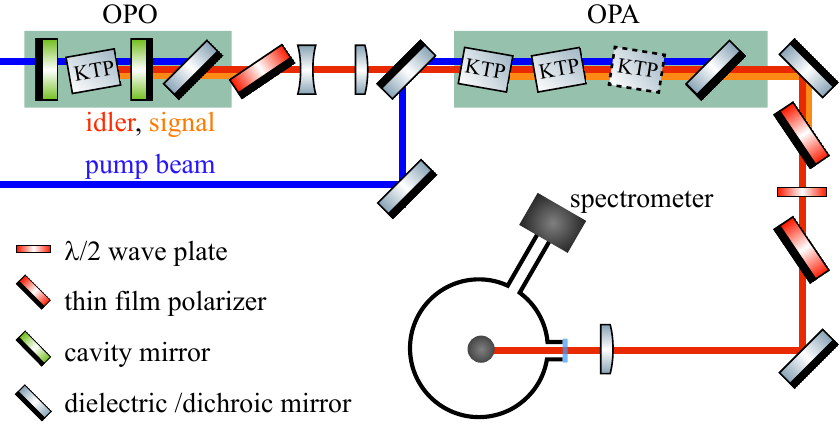}
		\caption{Schematic representation of the experimental setup. A master oscillator power amplifier (MOPA) setup, comprising an optical parametric oscillator (OPO) and an optical parametric amplifier (OPA), is pumped by a Nd:YAG laser (blue line). The signal beam is separated via polarization optics and the idler beam ($\lambda=$2.17\,\textmu m) is focused onto tin microdroplets within a vacuum chamber. EUV emission is captured by a transmission grating spectrometer positioned at \SI{60}{\degree} with respect to the laser axis. An additional, third KTP crystal (dashed outline) was used in the OPA in a subset of the experiments.}
		\label{fig:temp1}
	\end{figure}

	In the first set of experiments micrometer-sized liquid tin droplets are irradiated with high-intensity 2-\textmu m-wavelength laser pulses produced in a master oscillator power amplifier (MOPA). Following the work of Arisholm \textit{et al.}\,\cite{Arisholm2004}, the MOPA consists of a singly resonant optical parametric oscillator (OPO) in collinear alignment followed by an optical parametric amplifier (OPA). The latter comprises two 18-mm long KTP crystals operated in type II phase matching. The setup (see Fig.\,\ref{fig:temp1}) is pumped by a seeded Nd:YAG laser with a spatially flat-top and a temporally Gaussian profile of 10\,ns (FWHM). The OPO is pumped with 18\,mJ within a 1.5-mm diameter beam resulting in an idler beam energy of 1.8\,mJ at a wavelength of 2.17\,\textmu m. The OPO is operated slightly off its degeneracy point to minimize back conversion of signal and idler into the pump wavelength, a process which reduces beam quality of both beams. After the OPO the signal beam is removed and the idler beam expanded to 11\,mm and amplified in the OPA. Using  1.3\,J of pump energy within a beam diameter of 10\,mm, 260\,mJ of 2-\,\textmu m radiation are obtained, summing signal and idler pulse energies. The pulse duration of both beams after amplification is 4.3\,ns.

\begin{figure}[tb]
		\centering
		\includegraphics[scale=1]{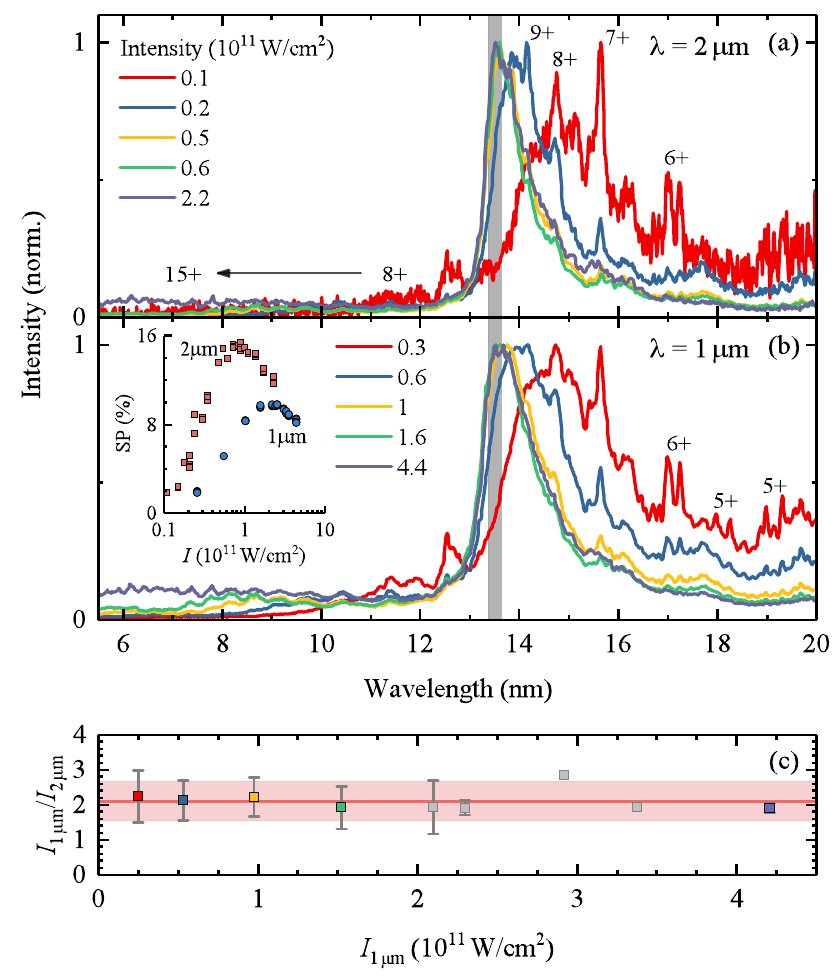}
		\caption{Spectra from tin droplet plasma observed at various intensities of the (a) 2-\textmu m beam (4.7\,ns pulse duration, 106\,\textmu m FWHM, 3 KTP crystals) and the (b) 1-\textmu m Nd:YAG beam (10\,ns pulse duration, 86\,\textmu m FWHM). The droplet size in both cases is 30\,\textmu m.
		(c) Ratio of the intensities of 1- and 2-\textmu m drive laser beams needed to obtain spectra with similar spectral features. The data points represent the average intensity ratios from data taken with four different laser spot sizes of the 2\,\textmu m laser beam of 65x88, 106, 152 and 194\,\textmu m (FWHM), respectively. The error bars indicate the standard deviation per measurement. The red line represents the average over all data points and the shaded band the standard deviation of the average.}
		\label{fig:temp2}
	\end{figure}

    For the experiments the signal beam is removed via polarization optics and the idler beam is solely used. The idler beam is focused onto several ten-micrometer-sized liquid tin droplets created via coalescence of even smaller microdroplets from a tin jet in a vacuum chamber that is kept at or below $10^{-6}$\,mbar. The diameter of the microdroplets is adjustable within a range from 16--51\,\textmu m. The focal spot is elliptical and has a size of 65x88\,\textmu m (FWHM) and laser intensities of up to \SI{2.1E11}{\intensity} are obtained on the tin droplets. Data taken with this 2-crystal setup is used for Sec.\,\ref{sec:depthscaling} and is part of the data in Fig.\ref{fig:temp2}(c).
    The intensity is defined as peak intensity in time and space and calculated to $I_L=(2\sqrt{2\ln{2}/2\pi})^3 E_L/ab t_p$ with laser energy $E_L$, FWHMs  $a$ and $b$ along the major and minor axis of the bivariate Gaussian and pulse duration $t_p$. The energy in the beam is adjusted by the combination of a half-waveplate and polarizer.
    The data displayed in Fig.\,\ref{fig:temp2}(a) was taken in a later experiment and after installation of a third crystal in the OPA which increased the energy in signal and idler combined to 360\,mJ while the pulse duration increased slightly to 4.7(3)\,ns (FWHM). 
    The produced beam has a symmetric focal spot and measurements are obtained for three focal spot sizes of 106, 152 and 194\,\textmu m (FWHM) that are obtained using lenses of different focal distance length. The data obtained in the first and this later experiment is combined in Fig.\,\ref{fig:temp2}(c).
    
    To enable a direct comparison with plasmas driven by 1-\textmu m wavelength laser pulses, light from the 1-\textmu m pump laser is redirected before entering the MOPA and is focused onto the tin droplets instead. Again, a combination of a half-wave plate and polarizer allows for adjustment of the beam energy. The focal spot has a symmetric Gaussian shape of 86\,\textmu m (FWHM).
    The EUV emission from the tin plasma is collected by a transmission grating spectrometer \cite{Bayraktar2016broadband} set up under a \SI{60}{\degree} angle with respect to the incoming laser beam. The spectrometer was operated with a 10,000\,lines/mm grating, a 25-\textmu m slit and without filter. The measured spectra are corrected for the grating's first and second order diffraction efficiency as well as for the quantum efficiency of the camera. The wavelength is calibrated in a separate experiment using atomic line emission from an aluminum plasma. 
    Spectral purity (SP), defined as the ratio of spectral energy in a 2\% bandwidth around 13.5\,nm to the total EUV energy, is used to characterize the EUV light source. All SP values provided are calculated with respect to the measured spectral range of 5.5--\SI{25.5}{nm}.

    \section{Scaling of spectral features with laser intensity and wavelength}
    \label{sec:spectralscaling}
    
    For defining development targets regarding power and pulse energy of future 2-\textmu m lasers for use in EUV light sources, it is particularly relevant to know the laser intensity needed to obtain a tin charge state balance optimal for the production of 13.5-nm light. In this section the laser intensity on the tin droplet is scanned and the optimal laser intensity determined as the value at which SP is highest, given that SP is the ultimate limit of CE as follows from energy conservation CE$<$SP/2 for isotropic emission\,\cite{Schupp2019}. To better understand the relevant plasma temperatures and densities, we study the ratio of 1- and 2-\textmu m  laser intensities at which plasmas of equal temperatures are established. Plasma temperature is experimentally established via the shape and amplitude of charge-state-specific spectral emission features \cite{Svendsen1994,Torretti2018,torretti2019spectral,Bouza2020}. These features are indicative of the plasma's charge state distribution which is predominantly dependent on plasma temperature\,\cite{Basko2015}. The experimental results are then compared to computer simulations using the radiation-hydrodynamic code RALEF-2D as well as to previous analytic work \cite{Basko2015}.

	\subsection{Spectral dependencies on drive laser intensity}

	In the experiments, first the idler beam from the MOPA is focused onto a 30-\textmu m diameter droplet and spectra are measured using the 106-\textmu m spot size at multiple intensities within a range of 0.1--\SI{2.2}{\intensity} (see Fig.\,\ref{fig:temp2}(a)). At the lowest laser intensity the plasma strongly emits around 14.5\,nm and distinct $4d$--$4f$ transitions in Sn$^{6+}$ are visible around 17\,nm \cite{Bouza2020}. Emission between 18 and 20\,nm can be mainly attributed to Sn$^{5+}$. At 15.7\,nm, a strong emission feature from  $4d$--$4f$ and $4p$--$4d$ transitions in Sn$^{7+}$ is visible. Going up this \q{ladder} of charge states, emission from $4d$--$4f$ and $4p$--$4d$ transitions in Sn$^{8+}$ is visible at 14.8\,nm and from Sn$^{9+}$ at 14.2\,nm. With increasing laser intensity the average charge state of the plasma increases and emission from Sn$^{10+}$ is evident in the 9.5--10-nm region\,\cite{Torretti2018}. Increasing laser intensity beyond \SI{E11}{\intensity}, the plasma strongly emits at 13.5\,nm. This emission originates from the $4d$--$4f$, $4d$--$5p$ and $4p$--$4d$ unresolved transitions arrays (UTAs) in Sn$^{8+}$ to Sn$^{14+}$ \cite{OSullivan2015, Churilov2006SnIX--SnXII}. With the strong emission at 13.5\,nm, charge state specific features become visible between 7 and 12\,nm. These features belong to the same Sn$^{8+}$ to Sn$^{14+}$ ions and here the $4d$--$5f$, $4d$--$6p$ and $4p$--$5s$ transitions contribute strongest \cite{Svendsen1994,Torretti2018}. With increasing laser intensity SP rises to values of 15\,\% at \SI{0.8e11}{\intensity} where charge state balance is optimal for in-band EUV emission, before reducing again at even higher intensity values (see inset in Fig.\,\ref{fig:temp2}(b)).

	Second, plasma is created using laser light of 1-\textmu m wavelength. Spectra for laser intensities within the range of 0.3--\SI{4.4e11}{\intensity} are shown in Fig.\,\ref{fig:temp2}(b). When compared to the 2-\textmu m drive-laser case the spectra show very similar shape, albeit at an apparent increased overall width. Again the same emission features of charge states Sn$^{5+}$ to Sn$^{9+}$ are visible at the lowest laser intensity but with somewhat less prominent emission features. This reduction in prominence is particularly noticeable for the peaks of charge states Sn$^{6+}$ and Sn$^{7+}$ (between 14 and 16\,nm). Further the Sn$^{9+}$ peak at 14.2\,nm is hardly visible (cf. \SI{0.2E11}{\intensity} in the 2-\textmu m case). The SP rises until it reaches values of 9.7\,\% around \SI{2E11}{\intensity} and subsequently decreases as the charge state balance becomes sub-optimal for emission of 13.5-nm light. The peak intensities used in this work agree well with previously published work, where the optimal SP was found at an intensity of \SI{1.4E11}{\intensity} using a temporally and spatially box-like laser profile to illuminate the tin droplets \cite{Schupp2019}. The higher intensity value found in this work is attributed to the fact that, because of their spatial extent, the droplets experience a slightly lower average intensity compared to the peak values stated.

	To obtain the sought-for laser-intensity ratio $I_{1\mu m}/I_{2\mu m}$, each spectrum of the 2-\textmu m laser case at intensity $I_{2\mu m}$ is matched to a spectrum of the 1-\textmu m case at intensity $I_{1\mu m}$ for which the resemblance of the relative amplitudes and shape of spectral features is best matching. As the spectral features are characteristic of individual tin charge states \cite{torretti2019spectral, Bouza2020} this comparison provides access to the scaling of the plasma's charge state distribution (and hence temperature) with laser wavelength. For each match of laser intensities the ratio  $I_{1\mu m}/I_{2\mu m}$ is calculated and plotted as a function of $I_{1\mu m}$ in Fig.\,\ref{fig:temp2}(c). The data points represent the average of comparisons made for multiple spot size conditions and for two droplet size conditions. In all cases spectra were compared to the ones taken with the 1-\textmu m wavelength laser beam size of 86\,\textmu m. More specifically, the comparison encompasses measurements with a 30-\textmu m diameter droplet for 2-\textmu m case beam sizes of 65x88, 106x106, 152x152 and 194x194\,\textmu m and on a 19-\textmu m diameter droplet for the  65x88-\textmu m beam. The red line shows the average  $I_{1\mu m}/I_{2\mu m}=2.1(6)$ of all measurements with the standard deviation (distribution width and not the error-on-the-mean) of the mean value as red shaded area. The depicted uncertainty is the standard deviation of the mean.

	\subsection{Theory and discussion}

     The temperature of a plasma can be expressed analytically if the equation of state (EOS) is sufficiently well known. The required EOS parameters will however depend on the location in the plasma where the laser light is absorbed. Two cases can be distinguished\,\cite{Basko2015}. Case I: absorption of laser light dominantly occurs close to the critical surface where the plasma's electron density equals the critical density ($n_e\approx n_{crit}\sim\lambda^{-2}$). This case is relevant for long wavelength laser light, e.g., from CO$_2$ lasers.  Case II: absorption is already significant in the underdense corona where the electron density is lower than the critical electron density. For laser absorption of 1- and 2-\textmu m beams, case II applies and the tin plasma temperature can be written as \cite{Basko2015}
	\begin{equation}
	    \label{eqn:Basko_T}
	    T \propto \left( \frac{1}{R\lambda^2} \right)^{-0.19} [I (1-\phi_r)]^{0.44},
	\end{equation}
	with laser wavelength $\lambda$, laser intensity $I$, radiative loss fraction $\phi_r$ of the plasma and characteristic radius of the sonic surface $R$, defined as the contour at which the ion velocity equals the local sound velocity. The numerical values for the powers -0.19 and 0.44 originate from the EOS\,\cite{Basko2015}. Differences in radiative losses of the plasmas are neglected in the following, as they may be small for similar density and temperature plasmas. The sonic surface $R$ is only slightly wavelength dependent and the small difference can be neglected. 
	From Eq.\,\eqref{eqn:Basko_T}, an intensity ratio $I_{i}/I_{j}=(\lambda_{j}/\lambda_{i})^{0.86}$ here $I_{1\mu m}/I_{2\mu m}=1.8$ is calculated for $\lambda=1$ and 2-\textmu m plasmas exhibiting equal plasma temperatures. The predicted ratio of 1.8 agrees well with the experimental one of 2.1(6) and well approximates a scaling with $\lambda^{-1}$.

	\begin{figure}
	    \centering
	    \includegraphics{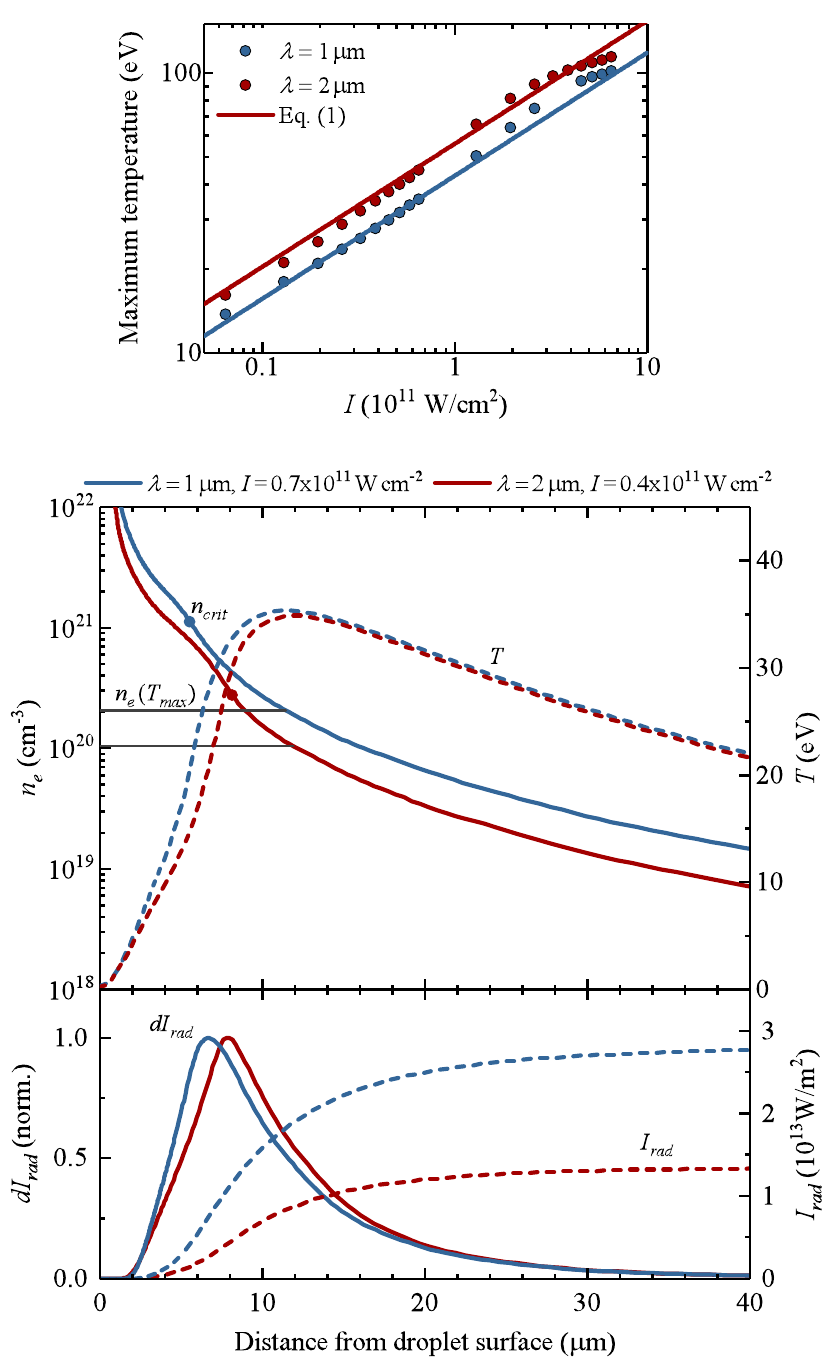}
	    \caption{Top: maximum temperature of a tin plasma for various laser intensities calculated with the two-dimensional-radiation transport code RALEF-2D. A 30-\textmu m diameter droplet is illuminated with temporally and spatially Gaussian-shaped laser pulses of wavelengths 1 and 2\,\textmu m.
	    Center: temperature and electron density lineout along the axis of the incoming laser beam.
	    Bottom: frequency-integrated local radiation field intensity $I_{rad}$ of the plasma and its normalized derivative $dI_{rad}$. The radiation field intensity is calculated from Eq\,\eqref{eq:radiation-field-intensity} using the density and temperature lineouts depicted in the center panel. For more detail see text.}
	    \label{fig:temp6} 
	\end{figure}

	Alongside this analytical approach, the radiation hydrodynamic code RALEF-2D \cite{Basko2017ralef} is used to determine the laser intensity ratio yielding equivalent plasma temperatures. RALEF-2D was developed to simulate laser plasma interaction and solves the equations of fluid dynamics in two dimensions (assuming cylindrical symmetry around the laser beam propagation axis) while including necessary physical mechanisms such as laser absorption, thermal conduction and radiation transport. The latter is needed for accurate predictions of a strongly radiating plasma, which is true for the current case. An extensive set of  simulations has been performed at conditions close to the experimental ones. A 30-\textmu m diameter droplet is irradiated by temporally and spatially Gaussian beams. The 1- and 2-\textmu m beams have pulse durations of 10 and 4.3\,ns (FWHM) and sizes of 80 and 100\,\textmu m (FWHM), respectively. Laser intensities in the range spanning \num{E10} to \SI{E12}{\intensity} are simulated. The plasma's peak temperature is plotted in Fig.\,\ref{fig:temp6}. For a given laser intensities the maximum temperature is consistently higher in the 2-\textmu m case. We note that the different pulse durations (10 vs. 4.3\,ns) have a minimal impact on temperature and density scales. The maximum temperatures are seen to follow Eq.\,\eqref{eqn:Basko_T} fitted as $T[\textrm{eV}] = a \, \lambda^{0.38}[\textrm{\textmu m}] \, I^{0.44}[\SI{E11}{\intensity}]$, where a common amplitude $a=43$ is determined by a global fit to all data. Eq.\,\eqref{eqn:Basko_T} captures the scaling of the peak plasma temperature over two decades in laser intensity. 
	
	Further shown in Fig.\,\ref{fig:temp6} are temperature and electron density lineouts along the laser axis away from the droplet at intensities relevant for the efficient emission of EUV light. The intensity of the 1- and 2-\textmu m cases were chosen to have nearly identical peak electron temperature. This temperature strongly increases with distance from the droplet surface and peaks around 11\,\textmu m from the droplet surface before it reduces again at larger distances. The maximum temperature is obtained at a factor of 2.0 lower density in the 2-\textmu m case. The point of highest temperature is much closer to the critical density in the 2-\textmu m case indicating that the absorption of laser light occurs closer to critical density while the conditions for laser absorption of case II are still met. Following Ref.\,\cite{Basko2015}, and references therein, the scaling of the relevant plasma electron density with wavelength can also be approximated invoking a constant absorbed fraction of the laser light, $k_L R$\,=\,constant. Inserting the Kramers' absorption coefficient for the laser radiation $k_L$ we obtain \cite{Basko2015,Oster1961emission}

	\begin{equation}
	   \label{eq:Basko_coronal_abs_cond}
	    (R \lambda^2) \rho^2 \bar{z}^3 T^{-3/2} = \mathrm{constant},
	\end{equation}

	with the mass density $\rho$ and the plasma's average charge state $\bar{z}$. Considering that mass density $\rho$ and ion density $n_i$ follow the ratio of electron density and average charge state $\rho \sim n_i=n_e/\bar{z}$, where $\bar{z}\approx 22.5 T^{0.6}$ \cite{Basko2015}, it becomes clear that the ratio of the electron densities lineouts displayed well approximates the ratio of mass density between the two laser wavelength cases. All other factors remaining constant in Eq.\,\eqref{eq:Basko_coronal_abs_cond}, a reciprocal scaling of mass density $\rho$ and wavelength $\lambda$ becomes directly apparent. The difference in mass density can thus be attributed to the difference in absorptivity of the laser radiation from Kramers' law\cite{Kramers1923}. This inversely proportional scaling of density with wavelength is the root cause of the observed intensity ratio.
	
	The bottom panel of Fig.\,\ref{fig:temp6} shows the radiation field intensity $I_{rad}$ and its normalized derivative $dI_{rad}$. The frequency-integrated radiation field intensity is calculated from 
	
	\begin{equation}
	\label{eq:radiation-field-intensity}
	    I_{rad}(s) = I_0 e ^{-\int^{s}_{s0} \alpha (s')ds'}
	    +\int^{s}_{s0} \alpha (s') B(s') e^{-\int^{s}_{s'}  \alpha (s'')ds''} ds'
	\end{equation}
	
	with the Planck mean absorptivity $\alpha_p[\mathrm{m}^{-1}] = \num{3.3E-7} \cdot \rho[\mathrm{g/cm}^{3}] \cdot T^{-1}[\mathrm{eV}]$ using the temperature and electron density information in Fig.\,\ref{fig:temp6}.	For more information see Ref.\,\cite{Torretti2020prominent}. The normalized derivatives $dI_{rad}$ peak at 6.5 and 8\,\textmu m distance from the droplet surface for the 1- and 2-\textmu m cases, respectively. They show that the typical length scales of emission are similar in both wavelength cases. The point of largest change in radiation field intensity is located slightly closer to the droplet surface than the point of maximum temperature. The significantly higher density more than compensates for the drop in temperature. 
	The point of largest change in the radiation field intensity of the 1-\textmu m driven plasma occurs relatively far from the critical density, whereas in the 2-\textmu m driven plasma this point lies close to the critical density, an observation explained by the distances between the respective maximum temperatures and critical densities. The radiation field intensity at large distances from the droplet surface is approximately a factor of two higher in the 1-\textmu m case because of the factor of two higher (emitter) density compared to the 2-\textmu m case.

	\section{Scaling of optical depth}
	\label{sec:depthscaling}
	
	The scaling of mass density with drive laser wavelength $\rho \sim \lambda^{-1}$ at similar length scales, as established by our simulations, indicates that the optical depth of the plasma, being a product of atomic opacity, mass density and path length, should scale similarly. If optical depth indeed reduces proportionally with drive laser wavelength, the step to a 2-\textmu m laser system could be particularly beneficial. 
    In the following, we perform an analysis of the optical depth associated with the EUV spectra by varying plasma size following the work of Schupp \textit{et al.}\,\cite{Schupp2019b}. This is accomplished by irradiating droplets having diameters in the range 16--51\,\textmu m. 
    \newpage
	
	\subsection{Scaling of peak optical depth with droplet size \\ and drive laser wavelength: examples}
	
\begin{figure}[tb]
		\centering
		\includegraphics[scale=1]{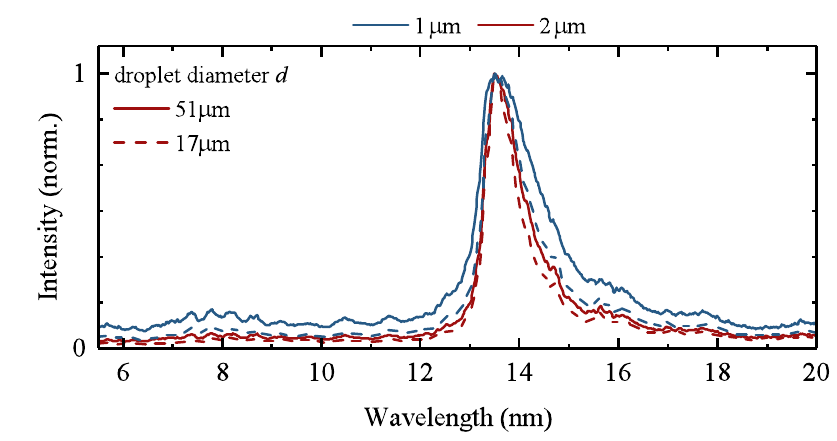}
		\caption{Spectral emission from tin plasmas produced with 1- and 2-\textmu m laser wavelength for small and large droplet diameters at laser intensities of 2.4 and  \SI{1.1e11}{\intensity}, respectively.}
		\label{fig:temp5}
	\end{figure}

	In our experiments the droplet diameter is changed in controlled steps from 16 to \SI{51}{\micro\meter} and a constant laser intensity is used for both laser wavelength cases.
	First, droplets are illuminated with 2-\textmu m laser light with an intensity of \SI{1.1E11}{\intensity}, close to optimal SP. The spot size is 65x88\,\textmu m. In Fig.\,\ref{fig:temp5} spectra for the smallest and largest droplet diameter are shown for both drive laser cases. With increasing droplet diameter the main emission feature at 13.5\,nm widens and more intense short wavelength radiation is emitted relative to the 13.5-nm peak. 
	Second, the same scan is repeated with 1-\textmu m laser light at \SI{2.4E11}{\intensity}, an intensity chosen based on the intensity ratio in Fig.\,\ref{fig:temp2}(c). Again, the main emission feature at 13.5\,nm widens with increasing droplet diameter and more intense short wavelength radiation is emitted relative to the 13.5-nm peak. For the 1-\textmu m driver these effects however are much stronger.

	In the following, the spectra are analyzed regarding their optical depth similar to the analysis in Ref.\,\onlinecite{Schupp2019b}. The wavelength-dependent optical depth $\tau_\lambda := \int \kappa_\lambda \rho dx$ is defined as the spatial integration over the product of the plasma's opacity $\kappa_\lambda$ and mass density $\rho$. In the instructive case of a one-dimensional plasma \cite{Bakshi2006} in local thermodynamic equilibrium (LTE), the spectral radiance is given by $L_\lambda = B_\lambda \left( 1 - e^{- \tau_{\lambda} } \right)$, where $B_\lambda$ is the Planck blackbody spectral radiance. We note that our high-density, strongly collisional 1- and 2-\textmu m driven plasmas are reasonably well approximated by LTE \cite{Torretti2020prominent}. At equal temperatures, and thus average charge state (recall $\bar{z}\approx 22.5 T^{0.6}$ \cite{Basko2015}), this equation enables each measured spectrum $\sim L_{\lambda,i}$ to be well approximated by any other spectrum $\sim L_{\lambda,j}$ when taking into account the ratio of the corresponding peak optical depths $a=\tau_{p,i}/\tau_{p,j}$ as a single parameter independent of wavelength (see Ref.\,\cite{Schupp2019b} and the Appendix for further details). Subsequently, if any peak optical depth $\tau_{p,j}$ is known in absolute terms, the optical depth of any other spectrum can be deduced. To be able to correct for systematic errors that could possibly occur for relatively low optical depth $\tau \lesssim 1$ we have suitably modified the equation used in Ref.\,\cite{Schupp2019b} as is detailed in the Appendix.
	
	As a reference spectrum, the spectrum measured at 1-\textmu m laser wavelength, 10-ns pulse duration and 30-\textmu m droplet size is chosen. The peak optical depth of this spectrum is determined by comparison of its 13.5-nm feature to opacity calculations in Ref.\,\cite{Torretti2020prominent}. More specifically, radiation transport is applied to the opacity spectrum calculated in Ref.\,\cite{Torretti2020prominent} for a here relevant mass density of $\rho=\SI{0.002}{g/cm^3}$ and electron temperature of $T_e=32$\,eV. The difference between radiation transported opacity spectrum and experimental spectrum is then minimized by changing the optical depth parameter $\tau_p$ in a least-square fit routine. This procedure leads to an absolute peak optical depth of $\tau_{0,p} = 4.5$ for our reference spectrum.

	\begin{figure}[tb]
		\centering
		\includegraphics[scale=1]{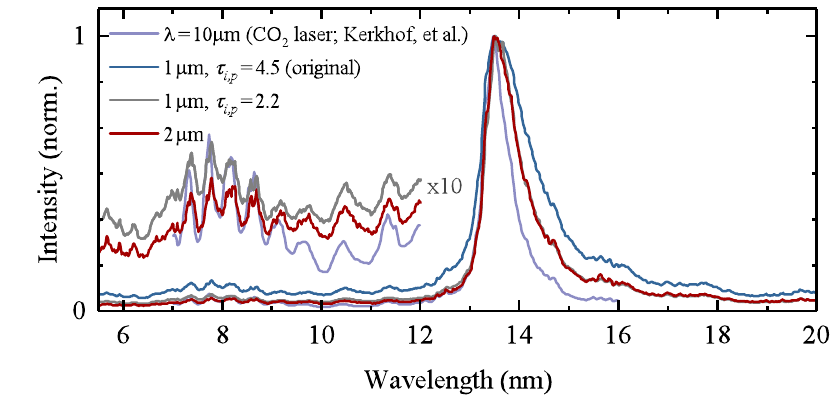}
		\caption{Spectrum produced with 2-\textmu m laser light (red line) compared to the radiation-transported reference spectrum for a peak optical depth value of $\tau_{p}=2.2$ (gray line, barely distinguishable from the red line). Reference and 2-\textmu m driven spectra were both obtained using a droplet diameter of 30\,\textmu m. Also shown is a spectrum obtained using a 10-\textmu m CO$_2$ laser \cite{Kerkhof2020} that represents the case of small optical depth.
		}
		\label{fig:temp3}
	\end{figure}
	
	Using Eq.\,\eqref{eqn:fit function_corrected} the peak optical depth $\tau_{i,p}$ of all spectra is fitted with respect to the reference spectrum. 
	As expected, inserting the relative optical depth obtained from the fits into Eq.\,\eqref{eqn:fit function} leads to an excellent reproduction of the main emission feature, as is shown in Fig.\,\ref{fig:temp3} for a typical example spectrum (30-\textmu m droplet with a 2-\textmu m driver). A further reasonable reproduction of the 7 to 12\,nm features is established with the 2-\textmu m driver outperforming the model spectrum with respect to the amount of radiation emitted out-of-band. Fig.\,\ref{fig:temp3} also shows a spectrum from an industrial plasma produced by a 10-\textmu m CO$_2$ driver which represents the limiting case of low optical depth. The step from a 1-\textmu m to a 2-\textmu m driver clearly significantly enhances the spectrum.

    \subsection{Scaling of peak optical depth with droplet size \\ and drive laser wavelength: all results}
	
\begin{figure}[!tb]
		\centering
		\includegraphics[scale=1]{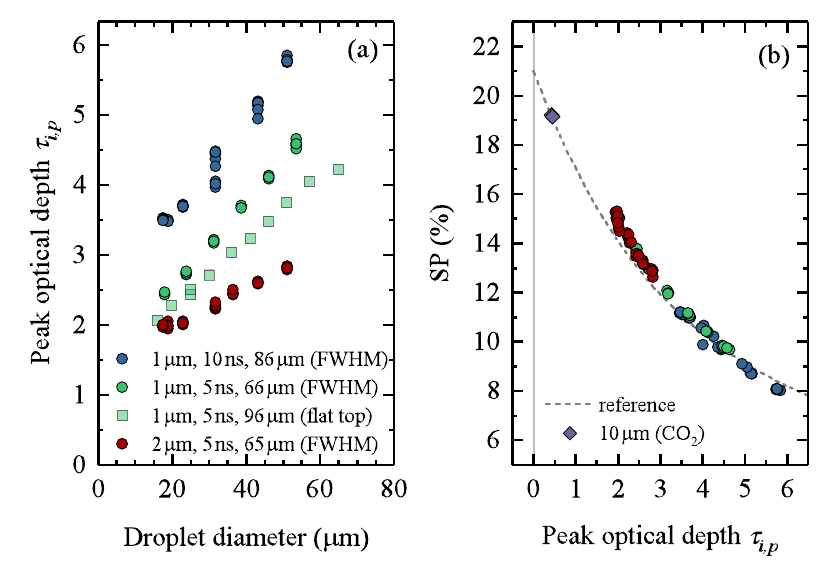}
		\caption{(a) Dependency of peak optical depth $ \tau_{i,p} $ on droplet diameter for 5- and 10-ns laser pulse duration at 1-\textmu m wavelength and for 4.3-ns pulse duration at  2-\textmu m wavelength. Circles indicate Gaussian spatial laser profile and boxes indicate a homogeneous 'flattop' laser beam profile.
		Peak optical depth is fitted with respect to the spectrum obtained at 1-\textmu m wavelength, 10-ns pulse duration and 30-\textmu m droplet diameter with optical depth of $\tau_{0,p}=4.5$.
		(b) Experimental values for spectral purity (SP) versus peak optical depth. The dashed line represents SP as calculated from the radiation-transported reference spectrum. The diamond symbol indicates the SP value of the radiation-transported reference spectrum for a peak optical depth value $\tau_{i,p}=0.4$, obtained from comparison of the reference spectrum with the emission of the CO$_2$-laser-driven plasma spectrum illustrated in Fig.\,\ref{fig:temp5}.
        }
		\label{fig:temp4}
	\end{figure}

	Having demonstrated the ability of the model function to accurately reproduce spectra from a single reference spectrum, we show in Fig.\,\ref{fig:temp4}(a) the fitted values for all spectra of the droplet size scans for 1- and 2-\textmu m laser wavelength. 
	In all cases the peak optical depth $ \tau_{i,p} $ appears to linearly increase with droplet diameter and to strongly depend on the laser wavelength. Indeed, the peak optical depth of the 2-\textmu m driven plasma lies roughly a factor of 2 below that of the 1-\textmu m one at largest droplet size, which may be expected from the lower plasma density (cf. Section\,\ref{sec:spectralscaling}). However, the 1-\textmu m results were obtained with 10-ns-long pulses and are here compared to the results from $\sim 5$-ns long, 2-\textmu m pulses, and optical depth is known to increase with pulse length \cite{Schupp2019,Schupp2019b}. To provide a comparison on more equal footing, we further compare in Fig.\,\ref{fig:temp4}(a) our results to previous data \cite{Schupp2019b}, obtained using a 1-\textmu m wavelength laser with a 5\,ns temporally box-shaped laser pulse. One of these data sets is taken with a spatially flattop laser profile of 96-\textmu m diameter \cite{Schupp2019,Schupp2019b} while the other one is taken with a Gaussian laser beam profile of 66\,\textmu m FWHM which more closely resembles the experimental conditions for the 2-\textmu m driver case. The spatial intensity distribution of the 1-\textmu m laser beam is seen to impact the effective optical depth (see also Ref.\,\cite{Schupp2019}). On comparison of the spectra for the 2- and 1-\textmu m cases at the most comparable temporal and spatial beam conditions, the clear reduction in peak optical depth parameter is maintained. This reduction, up to a factor 1.6 in optical depth becomes more pronounced at larger droplet diameters. The small deviation from the factor of $\sim$2 from the $\rho \sim \lambda^{-1}$ scaling may originate from differences in plasma length scales, plasma temperature, or from the finite laser intensity gradient over the plasma length scale. Nevertheless, a very significant reduction in optical depth of up to 40\,\% is demonstrated when using a 2-\textmu m laser to drive the plasma. \\
	
	With peak optical depth being the pertinent scaling parameter of 1- and 2-\textmu m driven tin plasmas the corresponding spectral purity of the emission spectrum is related to it in Fig.\,\ref{fig:temp4}(b). All experimental SP values, calculated over the range of 5.5--\SI{25.5}{nm}, collapse onto the gray dashed curve obtained by calculating the SP of the radiation-transported reference spectrum. The 2-\textmu m case is slightly offset towards higher SP values because of the reduced emission in the 7 to 12\,nm wavelength band compared to the radiation transported reference spectrum (cf. Fig.\,\ref{fig:temp3}) that is not captured by the model with the same accuracy as that of the main emission feature at 13.5\,nm. This difference between model and experiment may point to a small overestimation of the optical depth of the 2-\textmu m-laser-produced tin plasma, which would explain both the observed overestimation of the short-wavelength out-of-band emission by the model as well as the offset in Fig.\,\ref{fig:temp4}(b). This small overestimation of the optical depth may in turn be due to a broader charge state distribution in our measurements of the 2-\textmu m case caused by, e.g., laser intensity gradients or the slightly lower beam pointing stability compared to the 1-\textmu m case. This observation leads us to expect an even lower optical depth in the 2-\textmu m case and brings our scaling ratio even closer to the expected factor 2 from $~\lambda^{-1}$ scaling. More importantly, it indicates that there are further opportunities for narrowing the charge state distribution by providing a more homogeneous heating of the plasma in time and space. Such a narrowing of the charge state distribution around the optimum charge states Sn$^{11+}$--Sn$^{14+}$ would lead to further improvements of SP and thus CE.

\section{Conclusions}
	
	In conclusion, the effects of optical depth, plasma density, and laser intensity on the emission spectra of a 2-\textmu m-LPP source of tin microdroplets are investigated. The results are compared to the case of a 1-\textmu m driven plasma. It is found that the laser intensity required to maintain a common plasma temperature, scales approximately inversely with laser wavelength in going from 1- to 2-\textmu m drive laser, a result that will help defining development goals for future 2-\textmu m drive lasers for LPP light sources. 
	The reciprocal scaling with laser wavelength ($\sim\lambda^{-1}$) has its origin in Kramers' law of inverse Bremsstrahlung, the main laser absorption mechanism in the tin plasmas investigated. 
	Because of its reduced plasma density, the optical depth of the 2-\textmu m driven plasma is significantly reduced, allowing for efficient out-coupling of 13.5-nm radiation from the plasma even at larger plasma sizes. In future experiments it will be of interest to use large, pre-deformed targets and investigate the full CE potential of a 2-\textmu m source in a setting more similar to the current industrial one. Our results indicate that there are further opportunities for narrowing the charge state distribution by providing a more homogeneous heating of the plasma in time and space which would lead to further improvements of SP and thus CE. Looking further, it is of interest to experimentally investigate plasma generation using even longer-wavelength laser systems between 2 and 10\,\textmu m to find the mid-infrared wavelength optimally suited to drive EUV light sources at 13.5\,nm.\\

\section*{Acknowledgements}	
    We thank Mikhail M. Basko for providing us with the RALEF-2D code and his advise aiding the simulation work presented in the paper. Further we thank the authors of Ref.\,\cite{Kerkhof2020} for providing us with the data for the CO$_2$ spectrum shown in Fig.\,\ref{fig:temp3}.
	This work has been carried out at the Advanced Research Center for Nanolithography (ARCNL), a public-private partnership of the University of Amsterdam (UvA), the Vrije Universiteit Amsterdam (VU), the Netherlands Organisation for Scientific Research (NWO) and the semiconductor equipment manufacturer ASML.
	The used transmission grating spectrometer has been developed in the Industrial Focus Group XUV Optics at University of Twente, and supported by the FOM Valorisation Prize 2011 awarded to F. Bijkerk and NanoNextNL Valorization Grant awarded to M. Bayraktar in 2015.
	This project has received funding from European Research Council (ERC) Starting Grant number 802648 and is part of the VIDI research programme with project number 15697, which is financed by NWO.

	\section*{References}
	\bibliographystyle{apsrev4-2}

\begin{thebibliography}{41}%
\makeatletter
\providecommand \@ifxundefined [1]{%
 \@ifx{#1\undefined}
}%
\providecommand \@ifnum [1]{%
 \ifnum #1\expandafter \@firstoftwo
 \else \expandafter \@secondoftwo
 \fi
}%
\providecommand \@ifx [1]{%
 \ifx #1\expandafter \@firstoftwo
 \else \expandafter \@secondoftwo
 \fi
}%
\providecommand \natexlab [1]{#1}%
\providecommand \enquote  [1]{``#1''}%
\providecommand \bibnamefont  [1]{#1}%
\providecommand \bibfnamefont [1]{#1}%
\providecommand \citenamefont [1]{#1}%
\providecommand \href@noop [0]{\@secondoftwo}%
\providecommand \href [0]{\begingroup \@sanitize@url \@href}%
\providecommand \@href[1]{\@@startlink{#1}\@@href}%
\providecommand \@@href[1]{\endgroup#1\@@endlink}%
\providecommand \@sanitize@url [0]{\catcode `\\12\catcode `\$12\catcode
  `\&12\catcode `\#12\catcode `\^12\catcode `\_12\catcode `\%12\relax}%
\providecommand \@@startlink[1]{}%
\providecommand \@@endlink[0]{}%
\providecommand \url  [0]{\begingroup\@sanitize@url \@url }%
\providecommand \@url [1]{\endgroup\@href {#1}{\urlprefix }}%
\providecommand \urlprefix  [0]{URL }%
\providecommand \Eprint [0]{\href }%
\providecommand \doibase [0]{https://doi.org/}%
\providecommand \selectlanguage [0]{\@gobble}%
\providecommand \bibinfo  [0]{\@secondoftwo}%
\providecommand \bibfield  [0]{\@secondoftwo}%
\providecommand \translation [1]{[#1]}%
\providecommand \BibitemOpen [0]{}%
\providecommand \bibitemStop [0]{}%
\providecommand \bibitemNoStop [0]{.\EOS\space}%
\providecommand \EOS [0]{\spacefactor3000\relax}%
\providecommand \BibitemShut  [1]{\csname bibitem#1\endcsname}%
\let\auto@bib@innerbib\@empty
\bibitem [{\citenamefont {Versolato}(2019)}]{versolato2019physics}%
  \BibitemOpen
  \bibfield  {author} {\bibinfo {author} {\bibfnamefont {O.~O.}\ \bibnamefont
  {Versolato}},\ }\href {https://doi.org/10.1088/1361-6595/ab3302} {\bibfield
  {journal} {\bibinfo  {journal} {Plasma Sources Sci. T.}\ }\textbf {\bibinfo
  {volume} {28}},\ \bibinfo {pages} {083001} (\bibinfo {year}
  {2019})}\BibitemShut {NoStop}%
\bibitem [{\citenamefont {Purvis}\ \emph {et~al.}(2018)\citenamefont {Purvis},
  \citenamefont {Fomenkov}, \citenamefont {Schafgans}, \citenamefont {Vargas},
  \citenamefont {Rich}, \citenamefont {Tao}, \citenamefont {Rokitski},
  \citenamefont {Mulder}, \citenamefont {Buurman}, \citenamefont {Kats} \emph
  {et~al.}}]{Purvis2018industrialization}%
  \BibitemOpen
  \bibfield  {author} {\bibinfo {author} {\bibfnamefont {M.}~\bibnamefont
  {Purvis}}, \bibinfo {author} {\bibfnamefont {I.~V.}\ \bibnamefont
  {Fomenkov}}, \bibinfo {author} {\bibfnamefont {A.~A.}\ \bibnamefont
  {Schafgans}}, \bibinfo {author} {\bibfnamefont {M.}~\bibnamefont {Vargas}},
  \bibinfo {author} {\bibfnamefont {S.}~\bibnamefont {Rich}}, \bibinfo {author}
  {\bibfnamefont {Y.}~\bibnamefont {Tao}}, \bibinfo {author} {\bibfnamefont
  {S.~I.}\ \bibnamefont {Rokitski}}, \bibinfo {author} {\bibfnamefont
  {M.}~\bibnamefont {Mulder}}, \bibinfo {author} {\bibfnamefont
  {E.}~\bibnamefont {Buurman}}, \bibinfo {author} {\bibfnamefont
  {M.}~\bibnamefont {Kats}}, \emph {et~al.},\ }in\ \href
  {https://doi.org/10.1117/12.2305955} {\emph {\bibinfo {booktitle} {Extreme
  Ultraviolet (EUV) Lithography IX}}},\ Vol.\ \bibinfo {volume} {10583}\
  (\bibinfo {organization} {SPIE},\ \bibinfo {year} {2018})\ p.\ \bibinfo
  {pages} {1058327}\BibitemShut {NoStop}%
\bibitem [{\citenamefont {Moore}(2018)}]{Moore2018euv}%
  \BibitemOpen
  \bibfield  {author} {\bibinfo {author} {\bibfnamefont {S.~K.}\ \bibnamefont
  {Moore}},\ }\href {https://doi.org/10.1109/MSPEC.2018.8241736} {\bibfield
  {journal} {\bibinfo  {journal} {IEEE Spectrum}\ }\textbf {\bibinfo {volume}
  {55}},\ \bibinfo {pages} {46} (\bibinfo {year} {2018})}\BibitemShut {NoStop}%
\bibitem [{\citenamefont {Schafgans}\ \emph {et~al.}(2015)\citenamefont
  {Schafgans}, \citenamefont {Brown}, \citenamefont {Fomenkov}, \citenamefont
  {Sandstrom}, \citenamefont {Ershov}, \citenamefont {Vaschenko}, \citenamefont
  {Rafac}, \citenamefont {Purvis}, \citenamefont {Rokitski}, \citenamefont
  {Tao} \emph {et~al.}}]{Schafgans2015performance}%
  \BibitemOpen
  \bibfield  {author} {\bibinfo {author} {\bibfnamefont {A.~A.}\ \bibnamefont
  {Schafgans}}, \bibinfo {author} {\bibfnamefont {D.~J.}\ \bibnamefont
  {Brown}}, \bibinfo {author} {\bibfnamefont {I.~V.}\ \bibnamefont {Fomenkov}},
  \bibinfo {author} {\bibfnamefont {R.}~\bibnamefont {Sandstrom}}, \bibinfo
  {author} {\bibfnamefont {A.}~\bibnamefont {Ershov}}, \bibinfo {author}
  {\bibfnamefont {G.}~\bibnamefont {Vaschenko}}, \bibinfo {author}
  {\bibfnamefont {R.}~\bibnamefont {Rafac}}, \bibinfo {author} {\bibfnamefont
  {M.}~\bibnamefont {Purvis}}, \bibinfo {author} {\bibfnamefont
  {S.}~\bibnamefont {Rokitski}}, \bibinfo {author} {\bibfnamefont
  {Y.}~\bibnamefont {Tao}}, \emph {et~al.},\ }in\ \href
  {https://doi.org/10.1117/12.2087421} {\emph {\bibinfo {booktitle} {Extreme
  Ultraviolet (EUV) Lithography VI}}},\ Vol.\ \bibinfo {volume} {9422}\
  (\bibinfo {organization} {SPIE},\ \bibinfo {year} {2015})\ p.\ \bibinfo
  {pages} {94220B}\BibitemShut {NoStop}%
\bibitem [{\citenamefont {O'Sullivan}\ \emph {et~al.}(2015)\citenamefont
  {O'Sullivan}, \citenamefont {Li}, \citenamefont {D'Arcy}, \citenamefont
  {Dunne}, \citenamefont {Hayden}, \citenamefont {Kilbane}, \citenamefont
  {McCormack}, \citenamefont {Ohashi}, \citenamefont {O'Reilly}, \citenamefont
  {Sheridan}, \citenamefont {Sokell}, \citenamefont {Suzuki},\ and\
  \citenamefont {Higashiguchi}}]{OSullivan2015}%
  \BibitemOpen
  \bibfield  {author} {\bibinfo {author} {\bibfnamefont {G.}~\bibnamefont
  {O'Sullivan}}, \bibinfo {author} {\bibfnamefont {B.}~\bibnamefont {Li}},
  \bibinfo {author} {\bibfnamefont {R.}~\bibnamefont {D'Arcy}}, \bibinfo
  {author} {\bibfnamefont {P.}~\bibnamefont {Dunne}}, \bibinfo {author}
  {\bibfnamefont {P.}~\bibnamefont {Hayden}}, \bibinfo {author} {\bibfnamefont
  {D.}~\bibnamefont {Kilbane}}, \bibinfo {author} {\bibfnamefont
  {T.}~\bibnamefont {McCormack}}, \bibinfo {author} {\bibfnamefont
  {H.}~\bibnamefont {Ohashi}}, \bibinfo {author} {\bibfnamefont
  {F.}~\bibnamefont {O'Reilly}}, \bibinfo {author} {\bibfnamefont
  {P.}~\bibnamefont {Sheridan}}, \bibinfo {author} {\bibfnamefont
  {E.}~\bibnamefont {Sokell}}, \bibinfo {author} {\bibfnamefont
  {C.}~\bibnamefont {Suzuki}},\ and\ \bibinfo {author} {\bibfnamefont
  {T.}~\bibnamefont {Higashiguchi}},\ }\href
  {https://doi.org/10.1088/0953-4075/48/14/144025} {\bibfield  {journal}
  {\bibinfo  {journal} {J. Phys. B: At. Mol. Opt. Phys.}\ }\textbf {\bibinfo
  {volume} {48}},\ \bibinfo {pages} {144025} (\bibinfo {year}
  {2015})}\BibitemShut {NoStop}%
\bibitem [{\citenamefont {Banine}\ \emph {et~al.}(2011)\citenamefont {Banine},
  \citenamefont {Koshelev},\ and\ \citenamefont {Swinkels}}]{Banine2011}%
  \BibitemOpen
  \bibfield  {author} {\bibinfo {author} {\bibfnamefont {V.~Y.}\ \bibnamefont
  {Banine}}, \bibinfo {author} {\bibfnamefont {K.~N.}\ \bibnamefont
  {Koshelev}},\ and\ \bibinfo {author} {\bibfnamefont {G.~H. P.~M.}\
  \bibnamefont {Swinkels}},\ }\href
  {https://doi.org/10.1088/0022-3727/44/25/253001} {\bibfield  {journal}
  {\bibinfo  {journal} {J. Phys. D: Appl. Phys.}\ }\textbf {\bibinfo {volume}
  {44}},\ \bibinfo {pages} {253001} (\bibinfo {year} {2011})}\BibitemShut
  {NoStop}%
\bibitem [{\citenamefont {Benschop}\ \emph {et~al.}(2008)\citenamefont
  {Benschop}, \citenamefont {Banine}, \citenamefont {Lok},\ and\ \citenamefont
  {Loopstra}}]{Benschop2008}%
  \BibitemOpen
  \bibfield  {author} {\bibinfo {author} {\bibfnamefont {J.}~\bibnamefont
  {Benschop}}, \bibinfo {author} {\bibfnamefont {V.}~\bibnamefont {Banine}},
  \bibinfo {author} {\bibfnamefont {S.}~\bibnamefont {Lok}},\ and\ \bibinfo
  {author} {\bibfnamefont {E.}~\bibnamefont {Loopstra}},\ }\href
  {https://doi.org/10.1116/1.3010737} {\bibfield  {journal} {\bibinfo
  {journal} {J. Vac. Sci. Technol. B}\ }\textbf {\bibinfo {volume} {26}},\
  \bibinfo {pages} {2204} (\bibinfo {year} {2008})}\BibitemShut {NoStop}%
\bibitem [{\citenamefont {Azarov}\ and\ \citenamefont
  {Joshi}(1993)}]{Azarov1993}%
  \BibitemOpen
  \bibfield  {author} {\bibinfo {author} {\bibfnamefont {V.~I.}\ \bibnamefont
  {Azarov}}\ and\ \bibinfo {author} {\bibfnamefont {Y.~N.}\ \bibnamefont
  {Joshi}},\ }\href {https://doi.org/10.1088/0953-4075/26/20/011} {\bibfield
  {journal} {\bibinfo  {journal} {J. Phys. B: At. Mol. Opt. Phys.}\ }\textbf
  {\bibinfo {volume} {26}},\ \bibinfo {pages} {3495} (\bibinfo {year}
  {1993})}\BibitemShut {NoStop}%
\bibitem [{\citenamefont {Churilov}\ and\ \citenamefont
  {Ryabtsev}(2006{\natexlab{a}})}]{Churilov2006SnIX--SnXII}%
  \BibitemOpen
  \bibfield  {author} {\bibinfo {author} {\bibfnamefont {S.~S.}\ \bibnamefont
  {Churilov}}\ and\ \bibinfo {author} {\bibfnamefont {A.~N.}\ \bibnamefont
  {Ryabtsev}},\ }\href {https://doi.org/10.1088/0031-8949/73/6/014} {\bibfield
  {journal} {\bibinfo  {journal} {Phys. Scr.}\ }\textbf {\bibinfo {volume}
  {73}},\ \bibinfo {pages} {614} (\bibinfo {year}
  {2006}{\natexlab{a}})}\BibitemShut {NoStop}%
\bibitem [{\citenamefont {Churilov}\ and\ \citenamefont
  {Ryabtsev}(2006{\natexlab{b}})}]{Churilov2006SnVIII}%
  \BibitemOpen
  \bibfield  {author} {\bibinfo {author} {\bibfnamefont {S.~S.}\ \bibnamefont
  {Churilov}}\ and\ \bibinfo {author} {\bibfnamefont {A.~N.}\ \bibnamefont
  {Ryabtsev}},\ }\href {https://doi.org/10.1134/S0030400X06050043} {\bibfield
  {journal} {\bibinfo  {journal} {Opt. Spectrosc.}\ }\textbf {\bibinfo {volume}
  {100}},\ \bibinfo {pages} {660} (\bibinfo {year}
  {2006}{\natexlab{b}})}\BibitemShut {NoStop}%
\bibitem [{\citenamefont {Churilov}\ and\ \citenamefont
  {Ryabtsev}(2006{\natexlab{c}})}]{Churilov2006SnXIII--XV}%
  \BibitemOpen
  \bibfield  {author} {\bibinfo {author} {\bibfnamefont {S.~S.}\ \bibnamefont
  {Churilov}}\ and\ \bibinfo {author} {\bibfnamefont {A.~N.}\ \bibnamefont
  {Ryabtsev}},\ }\href {https://doi.org/10.1134/S0030400X06080017} {\bibfield
  {journal} {\bibinfo  {journal} {Opt. Spectrosc.}\ }\textbf {\bibinfo {volume}
  {101}},\ \bibinfo {pages} {169} (\bibinfo {year}
  {2006}{\natexlab{c}})}\BibitemShut {NoStop}%
\bibitem [{\citenamefont {Ryabtsev}\ \emph {et~al.}(2008)\citenamefont
  {Ryabtsev}, \citenamefont {Kononov},\ and\ \citenamefont
  {Churilov}}]{Ryabtsev2008SnXIV}%
  \BibitemOpen
  \bibfield  {author} {\bibinfo {author} {\bibfnamefont {A.~N.}\ \bibnamefont
  {Ryabtsev}}, \bibinfo {author} {\bibfnamefont {{\'E}.~Y.}\ \bibnamefont
  {Kononov}},\ and\ \bibinfo {author} {\bibfnamefont {S.~S.}\ \bibnamefont
  {Churilov}},\ }\href {https://doi.org/10.1134/S0030400X08120060} {\bibfield
  {journal} {\bibinfo  {journal} {Opt. Spectrosc.}\ }\textbf {\bibinfo {volume}
  {105}},\ \bibinfo {pages} {844} (\bibinfo {year} {2008})}\BibitemShut
  {NoStop}%
\bibitem [{\citenamefont {Tolstikhina}\ \emph {et~al.}(2006)\citenamefont
  {Tolstikhina}, \citenamefont {Churilov}, \citenamefont {Ryabtsev},\ and\
  \citenamefont {Koshelev}}]{Tolstikhina2006ATOMICDATA}%
  \BibitemOpen
  \bibfield  {author} {\bibinfo {author} {\bibfnamefont {I.~Y.}\ \bibnamefont
  {Tolstikhina}}, \bibinfo {author} {\bibfnamefont {S.~S.}\ \bibnamefont
  {Churilov}}, \bibinfo {author} {\bibfnamefont {A.~N.}\ \bibnamefont
  {Ryabtsev}},\ and\ \bibinfo {author} {\bibfnamefont {K.~N.}\ \bibnamefont
  {Koshelev}},\ }in\ \href@noop {} {\emph {\bibinfo {booktitle} {EUV sources
  for lithography}}},\ \bibinfo {editor} {edited by\ \bibinfo {editor}
  {\bibfnamefont {V.}~\bibnamefont {Bakshi}}}\ (\bibinfo  {publisher} {SPIE
  Press},\ \bibinfo {year} {2006})\ Chap.~\bibinfo {chapter} {4}, pp.\ \bibinfo
  {pages} {113--148}\BibitemShut {NoStop}%
\bibitem [{\citenamefont {D'Arcy}\ \emph {et~al.}(2009)\citenamefont {D'Arcy},
  \citenamefont {Ohashi}, \citenamefont {Suda}, \citenamefont {Tanuma},
  \citenamefont {Fujioka}, \citenamefont {Nishimura}, \citenamefont
  {Nishihara}, \citenamefont {Suzuki}, \citenamefont {Kato}, \citenamefont
  {Koike}, \citenamefont {White},\ and\ \citenamefont
  {O'Sullivan}}]{DArcy2009a}%
  \BibitemOpen
  \bibfield  {author} {\bibinfo {author} {\bibfnamefont {R.}~\bibnamefont
  {D'Arcy}}, \bibinfo {author} {\bibfnamefont {H.}~\bibnamefont {Ohashi}},
  \bibinfo {author} {\bibfnamefont {S.}~\bibnamefont {Suda}}, \bibinfo {author}
  {\bibfnamefont {H.}~\bibnamefont {Tanuma}}, \bibinfo {author} {\bibfnamefont
  {S.}~\bibnamefont {Fujioka}}, \bibinfo {author} {\bibfnamefont
  {H.}~\bibnamefont {Nishimura}}, \bibinfo {author} {\bibfnamefont
  {K.}~\bibnamefont {Nishihara}}, \bibinfo {author} {\bibfnamefont
  {C.}~\bibnamefont {Suzuki}}, \bibinfo {author} {\bibfnamefont
  {T.}~\bibnamefont {Kato}}, \bibinfo {author} {\bibfnamefont {F.}~\bibnamefont
  {Koike}}, \bibinfo {author} {\bibfnamefont {J.}~\bibnamefont {White}},\ and\
  \bibinfo {author} {\bibfnamefont {G.}~\bibnamefont {O'Sullivan}},\ }\href
  {https://doi.org/10.1103/PhysRevA.79.042509} {\bibfield  {journal} {\bibinfo
  {journal} {Phys. Rev. A}\ }\textbf {\bibinfo {volume} {79}},\ \bibinfo
  {pages} {042509} (\bibinfo {year} {2009})}\BibitemShut {NoStop}%
\bibitem [{\citenamefont {Ohashi}\ \emph {et~al.}(2010)\citenamefont {Ohashi},
  \citenamefont {Suda}, \citenamefont {Tanuma}, \citenamefont {Fujioka},
  \citenamefont {Nishimura}, \citenamefont {Sasaki},\ and\ \citenamefont
  {Nishihara}}]{Ohashi2010}%
  \BibitemOpen
  \bibfield  {author} {\bibinfo {author} {\bibfnamefont {H.}~\bibnamefont
  {Ohashi}}, \bibinfo {author} {\bibfnamefont {S.}~\bibnamefont {Suda}},
  \bibinfo {author} {\bibfnamefont {H.}~\bibnamefont {Tanuma}}, \bibinfo
  {author} {\bibfnamefont {S.}~\bibnamefont {Fujioka}}, \bibinfo {author}
  {\bibfnamefont {H.}~\bibnamefont {Nishimura}}, \bibinfo {author}
  {\bibfnamefont {A.}~\bibnamefont {Sasaki}},\ and\ \bibinfo {author}
  {\bibfnamefont {K.}~\bibnamefont {Nishihara}},\ }\href
  {https://doi.org/10.1088/0953-4075/43/6/065204} {\bibfield  {journal}
  {\bibinfo  {journal} {J. Phys. B: At. Mol. Opt. Phys.}\ }\textbf {\bibinfo
  {volume} {43}},\ \bibinfo {pages} {065204} (\bibinfo {year}
  {2010})}\BibitemShut {NoStop}%
\bibitem [{\citenamefont {Colgan}\ \emph {et~al.}(2017)\citenamefont {Colgan},
  \citenamefont {Kilcrease}, \citenamefont {Abdallah}, \citenamefont
  {Sherrill}, \citenamefont {Fontes}, \citenamefont {Hakel},\ and\
  \citenamefont {Armstrong}}]{Colgan2017}%
  \BibitemOpen
  \bibfield  {author} {\bibinfo {author} {\bibfnamefont {J.}~\bibnamefont
  {Colgan}}, \bibinfo {author} {\bibfnamefont {D.}~\bibnamefont {Kilcrease}},
  \bibinfo {author} {\bibfnamefont {J.}~\bibnamefont {Abdallah}}, \bibinfo
  {author} {\bibfnamefont {M.}~\bibnamefont {Sherrill}}, \bibinfo {author}
  {\bibfnamefont {C.}~\bibnamefont {Fontes}}, \bibinfo {author} {\bibfnamefont
  {P.}~\bibnamefont {Hakel}},\ and\ \bibinfo {author} {\bibfnamefont
  {G.}~\bibnamefont {Armstrong}},\ }\href
  {https://doi.org/http://dx.doi.org/10.1016/j.hedp.2017.03.009} {\bibfield
  {journal} {\bibinfo  {journal} {High Energy Density Phys.}\ }\textbf
  {\bibinfo {volume} {23}},\ \bibinfo {pages} {133 } (\bibinfo {year}
  {2017})}\BibitemShut {NoStop}%
\bibitem [{\citenamefont {Torretti}\ \emph {et~al.}(2017)\citenamefont
  {Torretti}, \citenamefont {Windberger}, \citenamefont {Ryabtsev},
  \citenamefont {Dobrodey}, \citenamefont {Bekker}, \citenamefont {Ubachs},
  \citenamefont {Hoekstra}, \citenamefont {Kahl}, \citenamefont {Berengut},
  \citenamefont {L\'opez-Urrutia},\ and\ \citenamefont
  {Versolato}}]{Torretti2017}%
  \BibitemOpen
  \bibfield  {author} {\bibinfo {author} {\bibfnamefont {F.}~\bibnamefont
  {Torretti}}, \bibinfo {author} {\bibfnamefont {A.}~\bibnamefont
  {Windberger}}, \bibinfo {author} {\bibfnamefont {A.}~\bibnamefont
  {Ryabtsev}}, \bibinfo {author} {\bibfnamefont {S.}~\bibnamefont {Dobrodey}},
  \bibinfo {author} {\bibfnamefont {H.}~\bibnamefont {Bekker}}, \bibinfo
  {author} {\bibfnamefont {W.}~\bibnamefont {Ubachs}}, \bibinfo {author}
  {\bibfnamefont {R.}~\bibnamefont {Hoekstra}}, \bibinfo {author}
  {\bibfnamefont {E.~V.}\ \bibnamefont {Kahl}}, \bibinfo {author}
  {\bibfnamefont {J.~C.}\ \bibnamefont {Berengut}}, \bibinfo {author}
  {\bibfnamefont {J.~R.~C.}\ \bibnamefont {L\'opez-Urrutia}},\ and\ \bibinfo
  {author} {\bibfnamefont {O.~O.}\ \bibnamefont {Versolato}},\ }\href
  {https://doi.org/10.1103/PhysRevA.95.042503} {\bibfield  {journal} {\bibinfo
  {journal} {Phys. Rev. A}\ }\textbf {\bibinfo {volume} {95}},\ \bibinfo
  {pages} {042503} (\bibinfo {year} {2017})}\BibitemShut {NoStop}%
\bibitem [{\citenamefont {Scheers}\ \emph {et~al.}(2020)\citenamefont
  {Scheers}, \citenamefont {Shah}, \citenamefont {Ryabtsev}, \citenamefont
  {Bekker}, \citenamefont {Torretti}, \citenamefont {Sheil}, \citenamefont
  {Czapski}, \citenamefont {Berengut}, \citenamefont {Ubachs}, \citenamefont
  {L\'opez-Urrutia}, \citenamefont {Hoekstra},\ and\ \citenamefont
  {Versolato}}]{Scheers2020}%
  \BibitemOpen
  \bibfield  {author} {\bibinfo {author} {\bibfnamefont {J.}~\bibnamefont
  {Scheers}}, \bibinfo {author} {\bibfnamefont {C.}~\bibnamefont {Shah}},
  \bibinfo {author} {\bibfnamefont {A.}~\bibnamefont {Ryabtsev}}, \bibinfo
  {author} {\bibfnamefont {H.}~\bibnamefont {Bekker}}, \bibinfo {author}
  {\bibfnamefont {F.}~\bibnamefont {Torretti}}, \bibinfo {author}
  {\bibfnamefont {J.}~\bibnamefont {Sheil}}, \bibinfo {author} {\bibfnamefont
  {D.~A.}\ \bibnamefont {Czapski}}, \bibinfo {author} {\bibfnamefont {J.~C.}\
  \bibnamefont {Berengut}}, \bibinfo {author} {\bibfnamefont {W.}~\bibnamefont
  {Ubachs}}, \bibinfo {author} {\bibfnamefont {J.~R.~C.}\ \bibnamefont
  {L\'opez-Urrutia}}, \bibinfo {author} {\bibfnamefont {R.}~\bibnamefont
  {Hoekstra}},\ and\ \bibinfo {author} {\bibfnamefont {O.~O.}\ \bibnamefont
  {Versolato}},\ }\href {https://doi.org/10.1103/PhysRevA.101.062511}
  {\bibfield  {journal} {\bibinfo  {journal} {Phys. Rev. A}\ }\textbf {\bibinfo
  {volume} {101}},\ \bibinfo {pages} {062511} (\bibinfo {year}
  {2020})}\BibitemShut {NoStop}%
\bibitem [{\citenamefont {Bajt}\ \emph {et~al.}(2002)\citenamefont {Bajt},
  \citenamefont {Alameda}, \citenamefont {Barbee}, \citenamefont {Clift},
  \citenamefont {Folta}, \citenamefont {Kaufmann},\ and\ \citenamefont
  {Spiller}}]{Bajt2002}%
  \BibitemOpen
  \bibfield  {author} {\bibinfo {author} {\bibfnamefont {S.}~\bibnamefont
  {Bajt}}, \bibinfo {author} {\bibfnamefont {J.~B.}\ \bibnamefont {Alameda}},
  \bibinfo {author} {\bibfnamefont {T.~W.}\ \bibnamefont {Barbee},
  \bibfnamefont {Jr.}}, \bibinfo {author} {\bibfnamefont {W.~M.}\ \bibnamefont
  {Clift}}, \bibinfo {author} {\bibfnamefont {J.~A.}\ \bibnamefont {Folta}},
  \bibinfo {author} {\bibfnamefont {B.}~\bibnamefont {Kaufmann}},\ and\
  \bibinfo {author} {\bibfnamefont {E.~A.}\ \bibnamefont {Spiller}},\ }\href
  {https://doi.org/10.1117/1.1489426} {\bibfield  {journal} {\bibinfo
  {journal} {Opt. Eng.}\ }\textbf {\bibinfo {volume} {41}},\ \bibinfo {pages}
  {1797} (\bibinfo {year} {2002})}\BibitemShut {NoStop}%
\bibitem [{\citenamefont {Huang}\ \emph {et~al.}(2017)\citenamefont {Huang},
  \citenamefont {Medvedev}, \citenamefont {van~de Kruijs}, \citenamefont
  {Yakshin}, \citenamefont {Louis},\ and\ \citenamefont {Bijkerk}}]{Huang2017}%
  \BibitemOpen
  \bibfield  {author} {\bibinfo {author} {\bibfnamefont {Q.}~\bibnamefont
  {Huang}}, \bibinfo {author} {\bibfnamefont {V.}~\bibnamefont {Medvedev}},
  \bibinfo {author} {\bibfnamefont {R.}~\bibnamefont {van~de Kruijs}}, \bibinfo
  {author} {\bibfnamefont {A.}~\bibnamefont {Yakshin}}, \bibinfo {author}
  {\bibfnamefont {E.}~\bibnamefont {Louis}},\ and\ \bibinfo {author}
  {\bibfnamefont {F.}~\bibnamefont {Bijkerk}},\ }\href
  {https://doi.org/10.1063/1.4978290} {\bibfield  {journal} {\bibinfo
  {journal} {Appl. Phys. Rev.}\ }\textbf {\bibinfo {volume} {4}},\ \bibinfo
  {pages} {011104} (\bibinfo {year} {2017})}\BibitemShut {NoStop}%
\bibitem [{\citenamefont {Danson}\ \emph {et~al.}(2019)\citenamefont {Danson},
  \citenamefont {Haefner}, \citenamefont {Bromage}, \citenamefont {Butcher},
  \citenamefont {Chanteloup}, \citenamefont {Chowdhury}, \citenamefont
  {Galvanauskas}, \citenamefont {Gizzi}, \citenamefont {Hein}, \citenamefont
  {Hillier} \emph {et~al.}}]{Danson2019petawatt}%
  \BibitemOpen
  \bibfield  {author} {\bibinfo {author} {\bibfnamefont {C.~N.}\ \bibnamefont
  {Danson}}, \bibinfo {author} {\bibfnamefont {C.}~\bibnamefont {Haefner}},
  \bibinfo {author} {\bibfnamefont {J.}~\bibnamefont {Bromage}}, \bibinfo
  {author} {\bibfnamefont {T.}~\bibnamefont {Butcher}}, \bibinfo {author}
  {\bibfnamefont {J.-C.~F.}\ \bibnamefont {Chanteloup}}, \bibinfo {author}
  {\bibfnamefont {E.~A.}\ \bibnamefont {Chowdhury}}, \bibinfo {author}
  {\bibfnamefont {A.}~\bibnamefont {Galvanauskas}}, \bibinfo {author}
  {\bibfnamefont {L.~A.}\ \bibnamefont {Gizzi}}, \bibinfo {author}
  {\bibfnamefont {J.}~\bibnamefont {Hein}}, \bibinfo {author} {\bibfnamefont
  {D.~I.}\ \bibnamefont {Hillier}}, \emph {et~al.},\ }\href
  {doi:10.1017/hpl.2019.36} {\bibfield  {journal} {\bibinfo  {journal} {High
  Power Laser Sci. Eng.}\ }\textbf {\bibinfo {volume} {7}} (\bibinfo {year}
  {2019})}\BibitemShut {NoStop}%
\bibitem [{\citenamefont {Sistrunk}\ \emph {et~al.}(2019)\citenamefont
  {Sistrunk}, \citenamefont {Alessi}, \citenamefont {Bayramian}, \citenamefont
  {Chesnut}, \citenamefont {Erlandson}, \citenamefont {Galvin}, \citenamefont
  {Gibson}, \citenamefont {Nguyen}, \citenamefont {Reagan}, \citenamefont
  {Schaffers}, \citenamefont {Siders}, \citenamefont {Spinka},\ and\
  \citenamefont {Haefner}}]{Sistrunk2019}%
  \BibitemOpen
  \bibfield  {author} {\bibinfo {author} {\bibfnamefont {E.}~\bibnamefont
  {Sistrunk}}, \bibinfo {author} {\bibfnamefont {D.~A.}\ \bibnamefont
  {Alessi}}, \bibinfo {author} {\bibfnamefont {A.}~\bibnamefont {Bayramian}},
  \bibinfo {author} {\bibfnamefont {K.}~\bibnamefont {Chesnut}}, \bibinfo
  {author} {\bibfnamefont {A.}~\bibnamefont {Erlandson}}, \bibinfo {author}
  {\bibfnamefont {T.~C.}\ \bibnamefont {Galvin}}, \bibinfo {author}
  {\bibfnamefont {D.}~\bibnamefont {Gibson}}, \bibinfo {author} {\bibfnamefont
  {H.}~\bibnamefont {Nguyen}}, \bibinfo {author} {\bibfnamefont
  {B.}~\bibnamefont {Reagan}}, \bibinfo {author} {\bibfnamefont
  {K.}~\bibnamefont {Schaffers}}, \bibinfo {author} {\bibfnamefont {C.~W.}\
  \bibnamefont {Siders}}, \bibinfo {author} {\bibfnamefont {T.}~\bibnamefont
  {Spinka}},\ and\ \bibinfo {author} {\bibfnamefont {C.}~\bibnamefont
  {Haefner}},\ }in\ \href {https://doi.org/10.1117/12.2525380} {\emph {\bibinfo
  {booktitle} {{Short-pulse High-energy Lasers and Ultrafast Optical
  Technologies}}}},\ Vol.\ \bibinfo {volume} {11034},\ \bibinfo {editor}
  {edited by\ \bibinfo {editor} {\bibfnamefont {P.}~\bibnamefont {Bakule}}\
  and\ \bibinfo {editor} {\bibfnamefont {C.~L.}\ \bibnamefont {Haefner}}},\
  \bibinfo {organization} {International Society for Optics and Photonics}\
  (\bibinfo  {publisher} {SPIE},\ \bibinfo {year} {2019})\ pp.\ \bibinfo
  {pages} {1 -- 8}\BibitemShut {NoStop}%
\bibitem [{\citenamefont {Siders}\ \emph {et~al.}(2020)\citenamefont {Siders},
  \citenamefont {Erlandson}, \citenamefont {Galvin}, \citenamefont {Frank},
  \citenamefont {Langer}, \citenamefont {Reagan}, \citenamefont {Scott},
  \citenamefont {Sistrunk},\ and\ \citenamefont {Spinka}}]{Siders2019euvlitho}%
  \BibitemOpen
  \bibfield  {author} {\bibinfo {author} {\bibfnamefont {C.~W.}\ \bibnamefont
  {Siders}}, \bibinfo {author} {\bibfnamefont {A.~C.}\ \bibnamefont
  {Erlandson}}, \bibinfo {author} {\bibfnamefont {T.~C.}\ \bibnamefont
  {Galvin}}, \bibinfo {author} {\bibfnamefont {H.}~\bibnamefont {Frank}},
  \bibinfo {author} {\bibfnamefont {S.}~\bibnamefont {Langer}}, \bibinfo
  {author} {\bibfnamefont {B.~A.}\ \bibnamefont {Reagan}}, \bibinfo {author}
  {\bibfnamefont {H.}~\bibnamefont {Scott}}, \bibinfo {author} {\bibfnamefont
  {E.~F.}\ \bibnamefont {Sistrunk}},\ and\ \bibinfo {author} {\bibfnamefont
  {T.~M.}\ \bibnamefont {Spinka}},\ }\href@noop {} {\emph {\bibinfo {title}
  {Efficient high power laser drivers for next-generation High Power EUV
  sources}}},\ \bibinfo {organization} {EUV Litho} (\bibinfo {year} {2019
  (accessed July 20, 2020)}),\ \bibinfo {note}
  {https://www.euvlitho.com/2019/S22.pdf}\BibitemShut {NoStop}%
\bibitem [{\citenamefont {Schupp}\ \emph {et~al.}(2019)\citenamefont {Schupp},
  \citenamefont {Torretti}, \citenamefont {Meijer}, \citenamefont {Bayraktar},
  \citenamefont {Sheil}, \citenamefont {Scheers}, \citenamefont {Kurilovich},
  \citenamefont {Bayerle}, \citenamefont {Schafgans}, \citenamefont {Purvis},
  \citenamefont {Eikema}, \citenamefont {Witte}, \citenamefont {Ubachs},
  \citenamefont {Hoekstra},\ and\ \citenamefont {Versolato}}]{Schupp2019b}%
  \BibitemOpen
  \bibfield  {author} {\bibinfo {author} {\bibfnamefont {R.}~\bibnamefont
  {Schupp}}, \bibinfo {author} {\bibfnamefont {F.}~\bibnamefont {Torretti}},
  \bibinfo {author} {\bibfnamefont {R.~A.}\ \bibnamefont {Meijer}}, \bibinfo
  {author} {\bibfnamefont {M.}~\bibnamefont {Bayraktar}}, \bibinfo {author}
  {\bibfnamefont {J.}~\bibnamefont {Sheil}}, \bibinfo {author} {\bibfnamefont
  {J.}~\bibnamefont {Scheers}}, \bibinfo {author} {\bibfnamefont
  {D.}~\bibnamefont {Kurilovich}}, \bibinfo {author} {\bibfnamefont
  {A.}~\bibnamefont {Bayerle}}, \bibinfo {author} {\bibfnamefont {A.~A.}\
  \bibnamefont {Schafgans}}, \bibinfo {author} {\bibfnamefont {M.}~\bibnamefont
  {Purvis}}, \bibinfo {author} {\bibfnamefont {K.~S.~E.}\ \bibnamefont
  {Eikema}}, \bibinfo {author} {\bibfnamefont {S.}~\bibnamefont {Witte}},
  \bibinfo {author} {\bibfnamefont {W.}~\bibnamefont {Ubachs}}, \bibinfo
  {author} {\bibfnamefont {R.}~\bibnamefont {Hoekstra}},\ and\ \bibinfo
  {author} {\bibfnamefont {O.~O.}\ \bibnamefont {Versolato}},\ }\href
  {https://doi.org/10.1063/1.5117504} {\bibfield  {journal} {\bibinfo
  {journal} {Appl. Phys. Lett.}\ }\textbf {\bibinfo {volume} {115}},\ \bibinfo
  {pages} {124101} (\bibinfo {year} {2019})}\BibitemShut {NoStop}%
\bibitem [{\citenamefont {Freeman}\ \emph {et~al.}(2012)\citenamefont
  {Freeman}, \citenamefont {Harilal}, \citenamefont {Verhoff}, \citenamefont
  {Hassanein},\ and\ \citenamefont {Rice}}]{Freeman2012laser}%
  \BibitemOpen
  \bibfield  {author} {\bibinfo {author} {\bibfnamefont {J.}~\bibnamefont
  {Freeman}}, \bibinfo {author} {\bibfnamefont {S.}~\bibnamefont {Harilal}},
  \bibinfo {author} {\bibfnamefont {B.}~\bibnamefont {Verhoff}}, \bibinfo
  {author} {\bibfnamefont {A.}~\bibnamefont {Hassanein}},\ and\ \bibinfo
  {author} {\bibfnamefont {B.}~\bibnamefont {Rice}},\ }\href
  {https://doi.org/10.1088/0963-0252/21/5/055003} {\bibfield  {journal}
  {\bibinfo  {journal} {Plasma Sources Sci. Technol.}\ }\textbf {\bibinfo
  {volume} {21}},\ \bibinfo {pages} {055003} (\bibinfo {year}
  {2012})}\BibitemShut {NoStop}%
\bibitem [{\citenamefont {Harilal}\ \emph {et~al.}(2011)\citenamefont
  {Harilal}, \citenamefont {Sizyuk}, \citenamefont {Hassanein}, \citenamefont
  {Campos}, \citenamefont {Hough},\ and\ \citenamefont
  {Sizyuk}}]{Harilal2011effect}%
  \BibitemOpen
  \bibfield  {author} {\bibinfo {author} {\bibfnamefont {S.}~\bibnamefont
  {Harilal}}, \bibinfo {author} {\bibfnamefont {T.}~\bibnamefont {Sizyuk}},
  \bibinfo {author} {\bibfnamefont {A.}~\bibnamefont {Hassanein}}, \bibinfo
  {author} {\bibfnamefont {D.}~\bibnamefont {Campos}}, \bibinfo {author}
  {\bibfnamefont {P.}~\bibnamefont {Hough}},\ and\ \bibinfo {author}
  {\bibfnamefont {V.}~\bibnamefont {Sizyuk}},\ }\href
  {https://doi.org/10.1063/1.3562143} {\bibfield  {journal} {\bibinfo
  {journal} {J. Appl. Phys.}\ }\textbf {\bibinfo {volume} {109}},\ \bibinfo
  {pages} {063306} (\bibinfo {year} {2011})}\BibitemShut {NoStop}%
\bibitem [{\citenamefont {Basko}\ \emph {et~al.}(2010)\citenamefont {Basko},
  \citenamefont {Maruhn},\ and\ \citenamefont
  {Tauschwitz}}]{Basko2010development}%
  \BibitemOpen
  \bibfield  {author} {\bibinfo {author} {\bibfnamefont {M.}~\bibnamefont
  {Basko}}, \bibinfo {author} {\bibfnamefont {J.}~\bibnamefont {Maruhn}},\ and\
  \bibinfo {author} {\bibfnamefont {A.}~\bibnamefont {Tauschwitz}},\
  }\href@noop {} {\bibfield  {journal} {\bibinfo  {journal} {GSI report}\
  }\textbf {\bibinfo {volume} {1}},\ \bibinfo {pages} {410} (\bibinfo {year}
  {2010})}\BibitemShut {NoStop}%
\bibitem [{\citenamefont {Arisholm}\ \emph {et~al.}(2004)\citenamefont
  {Arisholm}, \citenamefont {Nordseth},\ and\ \citenamefont
  {Rustad}}]{Arisholm2004}%
  \BibitemOpen
  \bibfield  {author} {\bibinfo {author} {\bibfnamefont {G.}~\bibnamefont
  {Arisholm}}, \bibinfo {author} {\bibfnamefont {{\O}.}~\bibnamefont
  {Nordseth}},\ and\ \bibinfo {author} {\bibfnamefont {G.}~\bibnamefont
  {Rustad}},\ }\href {https://doi.org/10.1364/OPEX.12.004189} {\bibfield
  {journal} {\bibinfo  {journal} {Opt. Express}\ }\textbf {\bibinfo {volume}
  {12}},\ \bibinfo {pages} {4189} (\bibinfo {year} {2004})}\BibitemShut
  {NoStop}%
\bibitem [{\citenamefont {Bayraktar}\ \emph {et~al.}(2016)\citenamefont
  {Bayraktar}, \citenamefont {Bastiaens}, \citenamefont {Bruineman},
  \citenamefont {Vratzov},\ and\ \citenamefont
  {Bijkerk}}]{Bayraktar2016broadband}%
  \BibitemOpen
  \bibfield  {author} {\bibinfo {author} {\bibfnamefont {M.}~\bibnamefont
  {Bayraktar}}, \bibinfo {author} {\bibfnamefont {H.~M.}\ \bibnamefont
  {Bastiaens}}, \bibinfo {author} {\bibfnamefont {C.}~\bibnamefont
  {Bruineman}}, \bibinfo {author} {\bibfnamefont {B.}~\bibnamefont {Vratzov}},\
  and\ \bibinfo {author} {\bibfnamefont {F.}~\bibnamefont {Bijkerk}},\ }\href
  {https://ris.utwente.nl/ws/files/6488314/nevac2016-bayraktar.pdf} {\bibfield
  {journal} {\bibinfo  {journal} {NEVAC blad}\ }\textbf {\bibinfo {volume}
  {54}},\ \bibinfo {pages} {14} (\bibinfo {year} {2016})}\BibitemShut {NoStop}%
\bibitem [{\citenamefont {{R. Schupp, F. Torretti, R. A. Meijer, M. Bayraktar,
  J. Scheers, D. Kurilovich, A. Bayerle, K. S. E. Eikema, S. Witte, W. Ubachs,
  R. Hoekstra and O. O. Versolato}}(2019)}]{Schupp2019}%
  \BibitemOpen
  \bibfield  {author} {\bibinfo {author} {\bibnamefont {{R. Schupp, F.
  Torretti, R. A. Meijer, M. Bayraktar, J. Scheers, D. Kurilovich, A. Bayerle,
  K. S. E. Eikema, S. Witte, W. Ubachs, R. Hoekstra and O. O. Versolato}}},\
  }\href {https://doi.org/https://doi.org/10.1103/PhysRevApplied.12.014010}
  {\bibfield  {journal} {\bibinfo  {journal} {Phys. Rev. Appl.}\ }\textbf
  {\bibinfo {volume} {12}},\ \bibinfo {pages} {014010} (\bibinfo {year}
  {2019})}\BibitemShut {NoStop}%
\bibitem [{\citenamefont {Svendsen}\ and\ \citenamefont
  {O'Sullivan}(1994)}]{Svendsen1994}%
  \BibitemOpen
  \bibfield  {author} {\bibinfo {author} {\bibfnamefont {W.}~\bibnamefont
  {Svendsen}}\ and\ \bibinfo {author} {\bibfnamefont {G.}~\bibnamefont
  {O'Sullivan}},\ }\href {https://doi.org/10.1103/PhysRevA.50.3710} {\bibfield
  {journal} {\bibinfo  {journal} {Phys. Rev. A}\ }\textbf {\bibinfo {volume}
  {50}},\ \bibinfo {pages} {3710} (\bibinfo {year} {1994})}\BibitemShut
  {NoStop}%
\bibitem [{\citenamefont {Torretti}\ \emph {et~al.}(2018)\citenamefont
  {Torretti}, \citenamefont {Schupp}, \citenamefont {Kurilovich}, \citenamefont
  {Bayerle}, \citenamefont {Scheers}, \citenamefont {Ubachs}, \citenamefont
  {Hoekstra},\ and\ \citenamefont {Versolato}}]{Torretti2018}%
  \BibitemOpen
  \bibfield  {author} {\bibinfo {author} {\bibfnamefont {F.}~\bibnamefont
  {Torretti}}, \bibinfo {author} {\bibfnamefont {R.}~\bibnamefont {Schupp}},
  \bibinfo {author} {\bibfnamefont {D.}~\bibnamefont {Kurilovich}}, \bibinfo
  {author} {\bibfnamefont {A.}~\bibnamefont {Bayerle}}, \bibinfo {author}
  {\bibfnamefont {J.}~\bibnamefont {Scheers}}, \bibinfo {author} {\bibfnamefont
  {W.}~\bibnamefont {Ubachs}}, \bibinfo {author} {\bibfnamefont
  {R.}~\bibnamefont {Hoekstra}},\ and\ \bibinfo {author} {\bibfnamefont
  {O.}~\bibnamefont {Versolato}},\ }\href
  {https://doi.org/10.1088/1361-6455/aaa593} {\bibfield  {journal} {\bibinfo
  {journal} {J. Phys. B: At. Mol. Opt. Phys.}\ }\textbf {\bibinfo {volume}
  {51}},\ \bibinfo {pages} {045005} (\bibinfo {year} {2018})}\BibitemShut
  {NoStop}%
\bibitem [{\citenamefont {Torretti}\ \emph {et~al.}(2019)\citenamefont
  {Torretti}, \citenamefont {Liu}, \citenamefont {Bayraktar}, \citenamefont
  {Scheers}, \citenamefont {Bouza}, \citenamefont {Ubachs}, \citenamefont
  {Hoekstra},\ and\ \citenamefont {Versolato}}]{torretti2019spectral}%
  \BibitemOpen
  \bibfield  {author} {\bibinfo {author} {\bibfnamefont {F.}~\bibnamefont
  {Torretti}}, \bibinfo {author} {\bibfnamefont {F.}~\bibnamefont {Liu}},
  \bibinfo {author} {\bibfnamefont {M.}~\bibnamefont {Bayraktar}}, \bibinfo
  {author} {\bibfnamefont {J.}~\bibnamefont {Scheers}}, \bibinfo {author}
  {\bibfnamefont {Z.}~\bibnamefont {Bouza}}, \bibinfo {author} {\bibfnamefont
  {W.}~\bibnamefont {Ubachs}}, \bibinfo {author} {\bibfnamefont
  {R.}~\bibnamefont {Hoekstra}},\ and\ \bibinfo {author} {\bibfnamefont
  {O.}~\bibnamefont {Versolato}},\ }\href
  {https://doi.org/10.1088/1361-6463/ab56d4} {\bibfield  {journal} {\bibinfo
  {journal} {J. Phys. D Appl. Phys.}\ }\textbf {\bibinfo {volume} {53}},\
  \bibinfo {pages} {055204} (\bibinfo {year} {2019})}\BibitemShut {NoStop}%
\bibitem [{\citenamefont {Bouza}\ \emph {et~al.}(2020)\citenamefont {Bouza},
  \citenamefont {Scheers}, \citenamefont {Ryabtsev}, \citenamefont {Schupp},
  \citenamefont {Behnke}, \citenamefont {Shah}, \citenamefont {Sheil},
  \citenamefont {Bayraktar}, \citenamefont {L{\'o}pez-Urrutia}, \citenamefont
  {Ubachs} \emph {et~al.}}]{Bouza2020}%
  \BibitemOpen
  \bibfield  {author} {\bibinfo {author} {\bibfnamefont {Z.}~\bibnamefont
  {Bouza}}, \bibinfo {author} {\bibfnamefont {J.}~\bibnamefont {Scheers}},
  \bibinfo {author} {\bibfnamefont {A.~N.}\ \bibnamefont {Ryabtsev}}, \bibinfo
  {author} {\bibfnamefont {R.}~\bibnamefont {Schupp}}, \bibinfo {author}
  {\bibfnamefont {L.}~\bibnamefont {Behnke}}, \bibinfo {author} {\bibfnamefont
  {C.}~\bibnamefont {Shah}}, \bibinfo {author} {\bibfnamefont {J.}~\bibnamefont
  {Sheil}}, \bibinfo {author} {\bibfnamefont {M.}~\bibnamefont {Bayraktar}},
  \bibinfo {author} {\bibfnamefont {J.~C.}\ \bibnamefont {L{\'o}pez-Urrutia}},
  \bibinfo {author} {\bibfnamefont {W.}~\bibnamefont {Ubachs}}, \emph
  {et~al.},\ }\href {https://doi.org/10.1088/1361-6455/aba3a8} {\bibfield
  {journal} {\bibinfo  {journal} {J. Phys. B}\ } (\bibinfo {year}
  {2020})}\BibitemShut {NoStop}%
\bibitem [{\citenamefont {Basko}\ \emph {et~al.}(2015)\citenamefont {Basko},
  \citenamefont {Novikov},\ and\ \citenamefont {Grushin}}]{Basko2015}%
  \BibitemOpen
  \bibfield  {author} {\bibinfo {author} {\bibfnamefont {M.~M.}\ \bibnamefont
  {Basko}}, \bibinfo {author} {\bibfnamefont {V.~G.}\ \bibnamefont {Novikov}},\
  and\ \bibinfo {author} {\bibfnamefont {A.~S.}\ \bibnamefont {Grushin}},\
  }\href {https://doi.org/10.1063/1.4921334} {\bibfield  {journal} {\bibinfo
  {journal} {Phys. Plasmas}\ }\textbf {\bibinfo {volume} {22}},\ \bibinfo
  {pages} {053111} (\bibinfo {year} {2015})}\BibitemShut {NoStop}%
\bibitem [{\citenamefont {Basko}\ \emph {et~al.}(2017)\citenamefont {Basko},
  \citenamefont {Maruhn},\ and\ \citenamefont {Tauschwitz}}]{Basko2017ralef}%
  \BibitemOpen
  \bibfield  {author} {\bibinfo {author} {\bibfnamefont {M.}~\bibnamefont
  {Basko}}, \bibinfo {author} {\bibfnamefont {J.}~\bibnamefont {Maruhn}},\ and\
  \bibinfo {author} {\bibfnamefont {A.}~\bibnamefont {Tauschwitz}},\ }\href
  {http://www.basko.net/mm/RALEF/ralef-main-report.pdf} {\bibfield  {journal}
  {\bibinfo  {journal} {Main Report}\ } (\bibinfo {year} {2017})}\BibitemShut
  {NoStop}%
\bibitem [{\citenamefont {Oster}(1961)}]{Oster1961emission}%
  \BibitemOpen
  \bibfield  {author} {\bibinfo {author} {\bibfnamefont {L.}~\bibnamefont
  {Oster}},\ }\href {https://doi.org/10.1103/RevModPhys.33.525} {\bibfield
  {journal} {\bibinfo  {journal} {Rev. Mod. Phys.}\ }\textbf {\bibinfo {volume}
  {33}},\ \bibinfo {pages} {525} (\bibinfo {year} {1961})}\BibitemShut
  {NoStop}%
\bibitem [{\citenamefont {Kramers}(1923)}]{Kramers1923}%
  \BibitemOpen
  \bibfield  {author} {\bibinfo {author} {\bibfnamefont {H.~A.}\ \bibnamefont
  {Kramers}},\ }\href {https://doi.org/10.1080/14786442308565244} {\bibfield
  {journal} {\bibinfo  {journal} {Philos. Mag.}\ }\textbf {\bibinfo {volume}
  {46}},\ \bibinfo {pages} {836} (\bibinfo {year} {1923})}\BibitemShut
  {NoStop}%
\bibitem [{\citenamefont {Torretti}\ \emph {et~al.}(2020)\citenamefont
  {Torretti}, \citenamefont {Sheil}, \citenamefont {Schupp}, \citenamefont
  {Basko}, \citenamefont {Bayraktar}, \citenamefont {Meijer}, \citenamefont
  {Witte}, \citenamefont {Ubachs}, \citenamefont {Hoekstra}, \citenamefont
  {Versolato} \emph {et~al.}}]{Torretti2020prominent}%
  \BibitemOpen
  \bibfield  {author} {\bibinfo {author} {\bibfnamefont {F.}~\bibnamefont
  {Torretti}}, \bibinfo {author} {\bibfnamefont {J.}~\bibnamefont {Sheil}},
  \bibinfo {author} {\bibfnamefont {R.}~\bibnamefont {Schupp}}, \bibinfo
  {author} {\bibfnamefont {M.}~\bibnamefont {Basko}}, \bibinfo {author}
  {\bibfnamefont {M.}~\bibnamefont {Bayraktar}}, \bibinfo {author}
  {\bibfnamefont {R.}~\bibnamefont {Meijer}}, \bibinfo {author} {\bibfnamefont
  {S.}~\bibnamefont {Witte}}, \bibinfo {author} {\bibfnamefont
  {W.}~\bibnamefont {Ubachs}}, \bibinfo {author} {\bibfnamefont
  {R.}~\bibnamefont {Hoekstra}}, \bibinfo {author} {\bibfnamefont
  {O.}~\bibnamefont {Versolato}}, \emph {et~al.},\ }\href
  {https://doi.org/10.1038/s41467-020-15678-y} {\bibfield  {journal} {\bibinfo
  {journal} {Nat. Commun.}\ }\textbf {\bibinfo {volume} {11}},\ \bibinfo
  {pages} {2334} (\bibinfo {year} {2020})}\BibitemShut {NoStop}%
\bibitem [{\citenamefont {Bakshi}\ \emph {et~al.}(2006)\citenamefont {Bakshi}
  \emph {et~al.}}]{Bakshi2006}%
  \BibitemOpen
  \bibfield  {author} {\bibinfo {author} {\bibfnamefont {V.}~\bibnamefont
  {Bakshi}} \emph {et~al.},\ }\href@noop {} {\emph {\bibinfo {title} {EUV
  sources for lithography}}},\ Vol.\ \bibinfo {volume} {149}\ (\bibinfo
  {publisher} {SPIE press Bellingham, Washington},\ \bibinfo {year}
  {2006})\BibitemShut {NoStop}%
\bibitem [{\citenamefont {van~de Kerkhof}\ \emph {et~al.}(2020)\citenamefont
  {van~de Kerkhof}, \citenamefont {Liu}, \citenamefont {Meeuwissen},
  \citenamefont {Zhang}, \citenamefont {de~Kruif}, \citenamefont {Davydova},
  \citenamefont {Schiffelers}, \citenamefont {Wählisch}, \citenamefont {van
  Setten}, \citenamefont {Varenkamp}, \citenamefont {Ricken}, \citenamefont
  {de~Winter}, \citenamefont {McNamara},\ and\ \citenamefont
  {Bayraktar}}]{Kerkhof2020}%
  \BibitemOpen
  \bibfield  {author} {\bibinfo {author} {\bibfnamefont {M.}~\bibnamefont
  {van~de Kerkhof}}, \bibinfo {author} {\bibfnamefont {F.}~\bibnamefont {Liu}},
  \bibinfo {author} {\bibfnamefont {M.}~\bibnamefont {Meeuwissen}}, \bibinfo
  {author} {\bibfnamefont {X.}~\bibnamefont {Zhang}}, \bibinfo {author}
  {\bibfnamefont {R.}~\bibnamefont {de~Kruif}}, \bibinfo {author}
  {\bibfnamefont {N.}~\bibnamefont {Davydova}}, \bibinfo {author}
  {\bibfnamefont {G.}~\bibnamefont {Schiffelers}}, \bibinfo {author}
  {\bibfnamefont {F.}~\bibnamefont {Wählisch}}, \bibinfo {author}
  {\bibfnamefont {E.}~\bibnamefont {van Setten}}, \bibinfo {author}
  {\bibfnamefont {W.}~\bibnamefont {Varenkamp}}, \bibinfo {author}
  {\bibfnamefont {K.}~\bibnamefont {Ricken}}, \bibinfo {author} {\bibfnamefont
  {L.}~\bibnamefont {de~Winter}}, \bibinfo {author} {\bibfnamefont
  {J.}~\bibnamefont {McNamara}},\ and\ \bibinfo {author} {\bibfnamefont
  {M.}~\bibnamefont {Bayraktar}},\ }in\ \href
  {https://doi.org/10.1117/12.2551021} {\emph {\bibinfo {booktitle} {Extreme
  Ultraviolet (EUV) Lithography XI}}},\ Vol.\ \bibinfo {volume} {11323},\
  \bibinfo {editor} {edited by\ \bibinfo {editor} {\bibfnamefont {N.~M.}\
  \bibnamefont {Felix}}\ and\ \bibinfo {editor} {\bibfnamefont
  {A.}~\bibnamefont {Lio}}},\ \bibinfo {organization} {International Society
  for Optics and Photonics}\ (\bibinfo  {publisher} {SPIE},\ \bibinfo {year}
  {2020})\ pp.\ \bibinfo {pages} {364 -- 379}\BibitemShut {NoStop}%
\end{thebibliography}%
%

\newpage
\begin{appendix}
    
    \section*{Appendix}
	\subsection*{Radiation transport model}

	To determine peak optical depth in this work, the recorded spectra are analyzed in a manner  similar to that presented in Ref.\,\onlinecite{Schupp2019b}. In the following, the method from  Ref.\,\onlinecite{Schupp2019b} is first outlined briefly and is subsequently generalized for use with plasmas that are optically thin. The wavelength-dependent optical depth $\tau_\lambda := \int \kappa_\lambda \rho dx$ is defined as the spatial integration over the product of the plasma's opacity $\kappa_\lambda$ and mass density density $\rho$. The spectral radiance $L_\lambda$ of an extended one-dimensional plasma can be calculated by means of its optical depth as \cite{Bakshi2006}
	
	\begin{equation}
	\label{eqn:radiation transport}
	L_\lambda = S_\lambda \left( 1 - e^{- \tau_{\lambda} } \right).
	\end{equation}
	In local thermodynamic equilibrium (LTE), where collisional processes drive atomic level populations, the source function $S_\lambda$ equals the Planck blackbody function $B_\lambda$. Rearranging Eq.\,\eqref{eqn:radiation transport}, the optical depth of the recorded plasma spectrum can be obtained from its relative spectral radiance ${L_\lambda}/{B_\lambda}$
	\begin{equation}
	\label{eqn:optical depth}
	\tau_\lambda = -\ln 
	\left(
	    1 - \frac{L_\lambda}{B_\lambda}
	\right).
	\end{equation}
    The optical depths of two plasmas of similar temperatures, but with modestly different densities and length scales, may differ (in first approximation) only by a single wavelength-independent multiplicative factor $a_i$, relating the plasmas' optical depths via $\tau_{\lambda,i} = a_i \, \tau_{\lambda,0}$. Here $\tau_{0}$ and $\tau_{i}$ are the two wavelength-dependent optical depths of the reference spectrum and any other spectrum $i$, respectively. The relative spectral radiances of these two plasmas can be related to each other via Eq.\,\eqref{eqn:optical depth}	
	\begin{equation}
	\label{eqn:fit function}
	\frac{L_{\lambda,i}}{B_{\lambda}} 
	= 1 - 
	\left( 
	    1 - 
	    \frac{L_{\lambda,0}}{B_{\lambda}} 
	\right)^{\tau_{i}/\tau_{0}}.
	\end{equation}
	In order to apply Eq.\,\eqref{eqn:fit function} to the spectra measured, the relative spectral radiance of the spectra must be known. 
	To obtain the relative spectral radiance, the ratio of observed spectrum $O_{\lambda}$ (meaning the spectrum as recorded with the spectrometer) and blackbody function is normalized to the peak value at 13.5-nm wavelength (subscript $p$) by replacing $L$ with $\Tilde{L}_{\lambda} = O_{\lambda} B_{p} / O_{p}$. The normalized ratio $\Tilde{L}_{\lambda}/B_{\lambda}$ is then multiplied by the amplitude factor $1-e^{-\tau_{p}}$ obtained from  Eq.\,\eqref{eqn:optical depth}
	\begin{equation}
	\label{eqn:fit function_corrected}
	\frac{\Tilde{L}_{\lambda,i}}{B_{\lambda}} = 
	\frac{
    	1 - 
    	\left( 
    	    1 - 
    	    \frac{\Tilde{L}_{\lambda,0}}{B_{\lambda}} 
    	    (1-e^{-\tau_{0,p}}) 
    	\right)^{\tau_{i,p}/\tau_{0,p}}
	}
    {
        1-e^{-\tau_{i,p}}
    }.
	\end{equation}
    Note that the wavelength-dependent optical depth values ($\tau_{0,\lambda}$) from Eq.\,\eqref{eqn:fit function} have been exchanged by their peak values ($\tau_{0,p}$). This generalized equation allows for determination of peak optical depth in optically thin plasmas in LTE if the peak optical depth of one of the spectra is known. In the current analysis the use of Eq.\,\eqref{eqn:fit function_corrected} results in optical depth values that are mostly very similar, but some of which are up to 25\,\% lower for the smallest optical depths cases ($\tau \sim 2$), than when using Eq.\,\eqref{eqn:fit function}. 	
    Using Eq.\,\eqref{eqn:fit function_corrected} the peak optical depths $\tau_{i,p}$ of all spectra are fitted with respect to a reference spectrum of known peak optical depth (see main text).

\end{appendix}

\end{document}